\documentclass[12pt]{article}

\usepackage{bbold}
\usepackage{float}
\usepackage{amsmath, amssymb, setspace,cite}
\usepackage{graphicx}

\vspace{0.6cm}
\date{\today} 

\begin{document}

\hfill {\tt CERN-PH-TH/2012-120}

\def\thefootnote{\fnsymbol{footnote}}

\begin{center}
\Large\bf\boldmath
\vspace*{1.cm} 
Supersymmetric constraints from $B_s \to \mu^+\mu^-$ and $B\to K^*\mu^+\mu^-$ observables\unboldmath
\end{center}
\vspace{0.6cm}

\begin{center}
F.~Mahmoudi$^{1,2,}$\footnote{Electronic address: mahmoudi@in2p3.fr} 
S. Neshatpour$^{2,}$\footnote{Electronic address: neshatpour@clermont.in2p3.fr} and
J. Orloff$^{2,}$\footnote{Electronic address: orloff@in2p3.fr}\\[0.4cm] 
\vspace{0.6cm}
{\sl $^1$ CERN Theory Division, Physics Department\\ CH-1211 Geneva 23, Switzerland}\\[0.4cm]
{\sl $^2$ Clermont Universit{\'e}, Universit\'e Blaise Pascal, CNRS/IN2P3,\\
LPC, BP 10448, 63000 Clermont-Ferrand, France}
\end{center}

\renewcommand{\thefootnote}{\arabic{footnote}}
\setcounter{footnote}{0}

\vspace{1.cm}
\begin{abstract}
We study the implications of the recent LHCb limit and results on $B_s\to\mu^+\mu^-$ and $B\to K^*\mu^+\mu^-$ observables in the constrained SUSY scenarios. After discussing the Standard Model predictions and carefully estimating the theoretical errors, we show the constraining power of these observables in CMSSM and NUHM. The latest limit on BR($B_s\to\mu^+\mu^-$), being very close to the SM prediction, constrains strongly the large $\tan\beta$ regime and we show that the various angular observables from $B\to K^*\mu^+\mu^-$ decay can provide complementary information in particular for moderate $\tan\beta$ values.
\end{abstract}

\newpage

\section{Introduction}
The rare decays $B_s\to\mu^+\mu^-$ and $B\to K^*\mu^+\mu^-$ are sensitive probes of new particles arising in the extensions of the Standard Model (SM) and in particular Supersymmetry (SUSY). The measurements of these decays provide important constraints on the masses of new particles which are too heavy to be produced directly. 

At large $\tan\beta$, the SUSY contributions to the decay $B_s \to \mu^+ \mu^-$ is dominated by the exchange of neutral Higgs bosons, and it has been emphasised in many works \cite{Choudhury:1998ze,Babu:1999hn,Huang:2000sm,Ellis:2005sc,Carena:2006ai,Ellis:2007ss,Mahmoudi:2007gd,Eriksson:2008cx,Golowich:2011cx,Akeroyd:2011kd} that this decay receives large enhancement, and very restrictive constraints can be obtained on the supersymmetric parameters. This decay is currently searched for by three LHC experiments: LHCb, CMS and ATLAS, and recently LHCb collaboration reported a very strong limit on the branching ratio of 4.5 $\times 10^{-9}$ \cite{Aaij:2012ac} which is only about 15\% larger than the SM prediction. This improved limit further constrains the SUSY parameter space.
A possible signal, although with a low significance is also reported by the CDF collaboration \cite{Aaltonen:2011fi}.

The decay $B\to K^*\mu^+\mu^-$ on the other hand provides a variety of complementary 
observables as it gives access to angular distributions in addition to the differential branching fraction.
Experimentally the exclusive $B\to K^*\mu^+\mu^-$ decay is easier to measure compared to the theoretically cleaner inclusive mode $B\to X_s\mu^+\mu^-$.
However from a theoretical point of view in the exclusive mode there are large uncertainties, which come mostly from the $B\to K$ form factors.
Within the QCD factorisation \cite{Beneke:2001at,Beneke:2004dp}, simplifications can be made on the form factor description and by looking into the rich phenomenology of the various 
kinematic distributions, observables that have smaller dependency on the form factors can be defined \cite{Kruger:2005ep,Egede:2008uy}.
These observables prove to be important tools to study extensions of the SM 
\cite{Ali:1999mm,Buchalla:2000sk,Kruger:2000zg,Feldmann:2002iw,Kruger:2005ep,Lunghi:2006hc,
Altmannshofer:2008dz,Bobeth:2008ij,Egede:2008uy,Egede:2009tp,Egede:2010zc,Bobeth:2010wg,Lunghi:2010tr,Bobeth:2011gi} and they can provide valuable information on different sectors of the theory. The full determination of the angular distributions of $B\to K^*\mu^+\mu^-$ constitutes a worthwhile challenge to the present and future experiments. Several angular observables have already been measured by Belle, Babar, CDF and LHCb. The most precise measurements come from the recent LHCb analyses with 1~fb$^{-1}$ of integrated luminosity \cite{LHCb-CONF-2012-008}.

In this work we study for the first time the implications of the recent measurements of $B\to K^*\mu^+\mu^-$ observables on constrained SUSY scenarios and update the constraints from BR($B_s\to\mu^+\mu^-$). 
Our numerical analysis is performed with SuperIso v3.3 \cite{Mahmoudi:2007vz,Mahmoudi:2008tp} and we study two constrained SUSY models: CMSSM and NUHM. To give some insight on the origin of these constraints, a mapping of the CMSSM parameter space into the Wilson coefficients subspace for the most relevant operators is also provided, which displays interesting partial correlations and hierarchies.

This paper is organised as follows: in section 2 we present a theoretical introduction to the
decays $B_s\to \mu^+\mu^-$ and $B\to K^*\mu^+\mu^-$, provide the SM predictions and estimate the errors. In section 3 we summarise the experimental results and section 4 contains our numerical analysis of the constraints on the SUSY models that are obtained from the recent LHCb results. Conclusions are contained in section 5.

\section{Observables, Inputs and Theoretical Uncertainties}
The effective Hamiltonian describing the $b \to s \ell^+ \ell^-$ transitions has the following generic structure\footnote{We neglect the doubly Cabibbo-suppressed contributions of order $V_{ub}V_{us}^*$.}\cite{Chetyrkin:1997gb,Yan:2000dc}:
\begin{equation}
{\cal H}_{\rm eff}  =  -\frac{4G_{F}}{\sqrt{2}} V_{tb} V_{ts}^{*} \, \Bigl[\,\sum_{i=1}^{6}
    C_{i} O_{i} +  \sum_{i=7}^{10} \Bigl( C_{i} O_{i} + C_{i}^{\prime} O_{i}^{\prime} \Bigr)\;+
    \sum_{i=1}^{2} \Bigl( C_{Q_i} Q_{i} + C_{Q_i}^{\prime} Q_{i}^{\prime} \Bigr) \Bigr]\;
\end{equation}
where $Q_1$ and $Q_2$ are the scalar and pseudo-scalar operators and the primed operators are chirality flipped compared to the non-primed operators.
Physics contributions at scales higher than $\mu$ are summarized in the so called Wilson coefficients $C^{(\prime)}_{i,Q_i}(\mu)$ 
(a typical choice for the scale is $m_b$ for the B decays). The Wilson coefficients include contributions from all particles heavier than 
$\mu_b = \mathcal{O}(m_b)$, in the SM these contributions are the top quark and the electroweak bosons and in BSM possible heavy NP particles are also summarized in the 
Wilson coefficients. The local operators $O_i^{(\prime)}$ and $Q_i^{(\prime)}$ include the long distance contributions 
from scales lower than $\mu_b$.
In the SM the primed and (pseudo-)scalar operators are either highly suppressed or absent.
The most relevant operators for our work are
\begin{align}
\label{physical_basis}
O_7 &= \frac{e}{(4\pi)^2} m_b (\overline{s} \sigma^{\mu\nu} P_R b) F_{\mu\nu} \;, &
O_8 &= \frac{g}{(4\pi)^2} m_b (\bar{s} \sigma^{\mu \nu} T^a P_R b) G_{\mu \nu}^a \;, 
\\ \nonumber
O_9 &=  \frac{e^2}{(4\pi)^2} (\overline{s} \gamma^\mu P_L b) (\bar{\ell} \gamma_\mu \ell) \;, &
O_{10} &=  \frac{e^2}{(4\pi)^2} (\overline{s} \gamma^\mu P_L b) (\bar{\ell} \gamma_\mu \gamma_5 \ell)\;, \\ \nonumber
Q_1 &= \frac{e^2}{(4\pi)^2} (\bar{s} P_R b)(\bar{\ell}\,\ell)\;, &
Q_2 &=  \frac{e^2}{(4\pi)^2} (\bar{s} P_R b)(\bar{\ell}\gamma_5 \ell)\;,
\end{align}
where $P_{R,L}=(1 \pm \gamma_5)/2$ denote the chiral projectors and $m_b$ is the $\overline{MS}$ mass of the $b$ quark.
A full list of the operators as well as the analytical expressions for all the Wilson coefficients can be found in \cite{Mahmoudi:2008tp}. The renormalisation group equations for $C_1$--$C_8$ are given in \cite{Czakon:2006ss}, for $C_{Q_1}$ and $C_{Q_2}$ in \cite{Choudhury:1998ze} and we calculated the running of $C_9$ which is given in Appendix~\ref{app:c}.

\label{sec:2}

\subsection{$B_s\to\mu^+\mu^-$}
A stringent 95\% C.L. limit on the untagged branching ratio
$BR_\textrm{untag}(B_s\to\mu^+\mu^-) < 4.5\times 10^{-9}$ has recently
been obtained by the LHCb collaboration \cite{Aaij:2012ac}.  Taking
into account the precise measurement \cite{LHCb-CONF-2012-002} of the
fractional decay width difference between the $B_s$ heavy and light
mass eigenstates,
$y_s\equiv\Delta\Gamma_{B_s}/(2\Gamma_{B_s})=0.088\pm 0.014$, this
limit on the untagged branching ratio can be translated into an even
stronger limit on the CP-averaged branching ratio
$BR(B_s\to\mu^+\mu^-)\equiv \frac12 BR(B_s^0\to\mu^+\mu^-)+\frac12 BR(\bar
B_s^0\to\mu^+\mu^-)$, reading \cite{DescotesGenon:2011pb,deBruyn:2012wj,deBruyn:2012wk}
\begin{equation}
  \label{eq:2}
  BR(B_s\to\mu^+\mu^-)=(1-y_s)
  BR_\textrm{untag} (B_s\to\mu^+\mu^-)<4.1\times 10^{-9}\textrm{ at 95\% C.L.}
\end{equation}

In terms of Wilson coefficients, this average branching ratio
is expressed as \cite{Bobeth:2001sq,Mahmoudi:2008tp}:
\begin{eqnarray}
  \label{eq:Bs2mm_formula}
BR(B_s\to\mu^+\mu^-)&=&\frac{G_F^2 \alpha^2}{64\pi^2}f_{B_s}^2
m_{B_s}^3 |V_{tb}V_{ts}^*|^2\tau_{B_s}\sqrt{1-\frac{4m_\mu^2}{m_{B_s}^2}}\\
&&\times\left\{\left(1-\frac{4m_\mu^2}{m_{B_s}^2}\right)
  |C_{Q_1}-C'_{Q_1}|^2+\left|(C_{Q_2}-C'_{Q_2})+2(C_{10}-C'_{10})\frac{m_\mu}{m_{B_s}}\right|^2\right\}\,.\nonumber  
\end{eqnarray}
In the Standard Model, only $C_{10}$ is non-vanishing and gets its
largest contributions from a $Z$ penguin top loop (75\%), and from a
charmed box diagram (24\%) . With the inputs of Table
\ref{tab:input}, $C_{10}=-4.21$, from which
$BR(B_s\to\mu^+\mu^-)|_{SM}=(3.53\pm 0.38)\times 10^{-9}$. The latest
experimental limit thus severely restraints the room for new physics, and
its proximity with the 2$\sigma$ upper value calls for a discussion of
the uncertainties in this SM prediction.

The main uncertainty comes from the $B_s$ decay constant $f_{B_s}$,
which has recently been re-evaluated by independent lattice QCD groups
of Table~\ref{tab:lattice}.
\begin{table}
 \begin{center}
\begin{tabular}{|ll|l|l|}
 \hline
Lattice QCD Group & Ref.& $f_{B_s}$& $f_B$\\\hline
ETMC-11& \cite{Dimopoulos:2011gx}&$232\pm10$ MeV &$195\pm12$ MeV\\
Fermilab-MILC-11& \cite{Bazavov:2011aa,Neil:2011ku}&$242\pm9.5$ MeV &$197\pm9$ MeV\\
HPQCD-12&\cite{Na:2012kp}&$227\pm10$ MeV &$191\pm9$ MeV\\
\hline
Our choice&&$234\pm10$ MeV &$194\pm10$ MeV\\
\hline
\end{tabular}
\caption{average of lattice QCD results used in this work.  \label{tab:lattice}}
\end{center}
\end{table}
Their 4.3\% uncertainties agree, as do their results within these
uncertainties, so that we have chosen an average of these three results
in what follows. This implies a 8.7\% uncertainty on the
branching ratio. 

Notice this range covers the recently published result $f_{B_s}=225\pm4\,\textrm{MeV}$
of McNeile et al. \cite{McNeile:2011ng}, whose lower value and striking
precision dominates any weighted average including it, like the one in
\cite{Davies:2012qf} ($227\pm 4$~MeV) proposed by one of the authors
of \cite{McNeile:2011ng}, or the one on \cite{Latticeaverages}:
$227\pm6$~MeV. The smallness of the extrapolation error in this
work raises a number of new questions, and we prefer to keep our naive
but more conservative average. This choice has however little effect
on the new physics applications we have in mind, as these depend
mostly on the lower end of that range.

Another potential source of uncertainty comes from the choice of
scale at which the fine structure constant is used in
Eq.~(\ref{eq:Bs2mm_formula}): there is a non-negligible 4\% difference between the running
$\overline{MS}$ couplings $\hat\alpha(m_b)=1/133$ and
$\hat\alpha(m_Z)=1/128$. If the first choice may seem natural, the
weak couplings involved in the top $Z$-penguin (or charmed box) are
closer to the weak scale, and do not run below it, as discussed in
Ref.~\cite{Bobeth:2003at}. We thus take that last value, as well as
$\sin^2 \hat\theta_W(m_Z)= 0.2312$ in the expression of Eq.~(\ref{eq:Bs2mm_formula}).
This may seem at odds with the conclusion of Ref.~\cite{Bobeth:2003at},
that choosing $\hat\alpha(m_b)$ minimises the EW corrections to $B\to
K^* \ell^+ \ell^-$. 
\newpage
\noindent However, the EW corrections to $C_{7,9}$, which dominate the
low $q_{ll}^2$ region of this last process, are opposite to the EW
corrections for $C_{10}$, which controls $B_s\to \mu^+ \mu^-$. Having made this
choice for the EM-coupling, we expect EW corrections to mostly absorb the
remaining scale dependence in $\hat\alpha_{EM}(\mu)$, leaving a small,
2\% uncertainty in the branching ratio.

The remaining theoretical uncertainties are smaller thanks to the NNLO
treatment of QCD corrections: increasing the low scale $\mu_b$ (or the
matching scale $\mu_W$) by a factor of 2 induces a $1.4\%$ (or
respectively 2\%) effect.

Finally, parametric uncertainties from the top mass (1.3\%), from the $B_s$
lifetime (1.8\%) and from the CKM element $V_{ts}$ (5\%), will reduce in
the future. Adding all these (small) errors in quadrature, we thus get
a Standard Model prediction assorted with an 11\% uncertainty:
\begin{equation}
  \label{eq:1}
BR(B_s\to \mu^+ \mu^-)
= (3.53\pm0.38) \times 10^{-9} \;.
\end{equation}
This value is compatible with recent SM predictions for this observable, {\it e.g.} by the CKMfitter group
  \cite{Charles:2011va}, for which the uncertainties are reduced by the use of other flavour information.

\label{sec:Bsmumu}

\subsection{$B\to K^*\mu^+\mu^-$}
\begin{table}
\begin{center}
\footnotesize{\begin{tabular}{|lr|lr|}\hline
$m_B=5.27950$ GeV                         & \cite{Nakamura:2010}     &        $m_{B_s} = 5.3663 $ GeV& \cite{Nakamura:2010}                            \\
$m_{K^*}=0.89594$ GeV                     & \cite{Nakamura:2010}     & $|V_{tb}V_{ts}^*|=0.0403 ^{+0.0011}_{-0.0007}$         & \cite{Nakamura:2010}          \\ \hline
$m_b^{\overline{MS}}(m_b)=4.19 ^{+0.18}_{-0.06}$ GeV & \cite{Nakamura:2010}     & $m_c^{\overline{MS}}(m_c)=1.29 ^{+0.05}_{-0.11}$ GeV   & \cite{Nakamura:2010}\\ 
$m_t^{pole}=172.9 \pm0.6 \pm0.9$ GeV       & \cite{Nakamura:2010}     &$m_{\mu}=0.105658$ GeV                    & \cite{Nakamura:2010} \\ \hline  
$\alpha_s(M_Z)=0.1184 \pm 0.0007$         & \cite{Nakamura:2010}     & $\hat \alpha_{em}(M_Z)=1/127.916 $                     & \cite{Nakamura:2010}          \\ 
$\alpha_s(\mu_b)=0.2161$                  &                          &$\hat\alpha_{em}(m_b)=1/133$                               &           \\ 
$\sin^2\hat\theta_W(M_Z)=0.23116(13)$     & \cite{Nakamura:2010}&$G_F/(\hbar c)^3=1.16637(1)\;\textrm{GeV}^{-2}$& \cite{Nakamura:2010}\\ \hline
$ f_B=194 \pm 10$ MeV                          & Table~\ref{tab:lattice}& $\tau_B=1.519 \pm0.007$ ps                             & \cite{Nakamura:2010}          \\
$ f_{B_s} = 234 \pm 10 {\rm MeV}$ & Table~\ref{tab:lattice} & $ \tau_{B_s} = 1.472 \pm 0.026\ {\rm ps}    $ & \cite{Nakamura:2010} \\
 \hline
$f_{K^*,\perp}$(1 GeV)$=0.185 \pm0.009$ GeV  & \cite{Ball:2007}         & $f_{K^*,\parallel}=0.220 \pm0.005$ GeV                 & \cite{Ball:2007}              \\
$a_{1,\perp}$(1 GeV)$=0.10\pm0.07$          & \cite{Ball:2004rg}       & $a_{1,\parallel}$(1 GeV)$=0.10 \pm0.07$                   & \cite{Ball:2004rg}              \\
$a_{2,\perp}$(1 GeV)$=0.13 \pm0.08$          & \cite{Ball:2004rg}       & $a_{2,\parallel}$(1 GeV)$=0.09 \pm0.05$                   & \cite{Ball:2004rg}              \\
$V^{B_q\to K^*}(0) = 0.411 \pm 0.046$     & \cite{Ball:2004rg}       & $A_1^{B_q\to K^*}(0) = 0.292 \pm 0.038$                & \cite{Ball:2004rg}            \\
$\lambda_{B,+}$(1 GeV)$=0.46 \pm 0.11$ GeV   & \cite{Ball:2006nr}       & $A_2^{B_q\to K^*}(0) = 0.259 \pm 0.036$                & \cite{Ball:2004rg}            \\ \hline
$\mu_b=m_b^{pole}$                        &                          & $\mu_0=2 M_W$                                          &                               \\ 
$\mu_f=\sqrt{0.5 \times \mu_b}$ GeV       & \cite{Beneke:2004dp}     &                                                        &                               \\ \hline
\end{tabular}}
\caption{Input parameters used in this work.  \label{tab:input}}
\end{center}
\end{table}

Considering the $\bar K^*$ meson to be on-shell, the differential decay distribution of the $\bar B ^0 \to \bar K ^*(\to K^- \pi^+ ) \ell^+ \ell^-$ decay 
can be written in terms of three angles $\theta_\ell$, $\theta_{K^*}$, $\phi$ and 
the invariant dilepton mass squared ($q^2$) \cite{Kruger:2005ep,Altmannshofer:2008dz}:

\begin{equation}\label{eq:diffAD}
  d^4\Gamma = \frac{9}{32\pi} J(q^2, \theta_l, \theta_{K^*}, \phi)\, dq^2\, d\cos\theta_l\, d\cos\theta_{K^*}\, d\phi \;.
\end{equation}

\newpage
In the above equation $\theta_\ell$ is the angle between $\ell^-$ and $\bar B^0$ in the rest frame of the dilepton, 
 $\theta_{K^*}$ is the angle between $K^-$ and $\bar B^0$ in the $\bar K^*$ rest frame and 
$\phi$ is the angle between the normals of the $\ell^+ \ell^-$ plane and the $K^- \pi^+$ in the $\bar B^0$ rest frame. 
The angular dependence of $J(q^2, \theta_l, \theta_{K^*}, \phi)$ can be written as 

\begin{align}\label{eq:J}
  J(q^2, \theta_\ell, \theta_{K^*}, \phi)& = J_1^s \sin^2\theta_{K^*} + J_1^c \cos^2\theta_{K^*}
      + (J_2^s \sin^2\theta_{K^*} + J_2^c \cos^2\theta_{K^*}) \cos 2\theta_\ell
\nonumber \\       
    & + J_3 \sin^2\theta_{K^*} \sin^2\theta_\ell \cos 2\phi 
      + J_4 \sin 2\theta_{K^*} \sin 2\theta_\ell \cos\phi 
      + J_5 \sin 2\theta_{K^*} \sin\theta_\ell \cos\phi
\nonumber \\      
    & + (J_6^s \sin^2\theta_{K^*} + J_6^c \cos^2\theta_{K^*})\cos\theta_\ell 
      + J_7 \sin 2\theta_{K^*} \sin\theta_\ell \sin\phi
\nonumber \\ 
    & + J_8 \sin 2\theta_{K^*} \sin 2\theta_\ell \sin\phi
      + J_9 \sin^2\theta_{K^*} \sin^2\theta_\ell \sin 2\phi \;. 
\end{align}

The angular coefficients $J_i$ (see Appendix \ref{app:a}), are functions of $q^2$ and can be described in terms of eight transversity amplitudes 
$A_{\perp}^{L,R}$, $A_{\parallel}^{L,R}$, $A_{0}^{L,R}$, $A_t$, $A_S$.
The transversity amplitudes up to corrections of ${\cal O}(\alpha_s)$ can be expressed in terms of seven independent form factors, $A_{0,1,2}, T_{1,2,3}$ and $V$.
Since these form factors are hadronic quantities they require non-perturbative calculations and hence are a main source of uncertainty in the exclusive mode.
But even if the form factors were known precisely the $\bar B ^0 \to \bar K ^* \ell^+ \ell^-$ decay 
would still need corrections emerging from non-factorisable effects which are related to the current-current operators $O_1$ and $O_2$,
the QCD penguin operators $O_3$-$O_6$ and the chromomagnetic operator $O_8$. The non-factorisable corrections contribute to the decay amplitude by producing a virtual
photon which decays into a lepton pair. 
When the $\bar K^*$ has a large energy ($q^2$ is small) and the decaying hadron ($\bar B^0$) is heavy, 
the non-factorisable corrections can be computed in the QCD factorisation framework \cite{Beneke:2001at,Beneke:2004dp}.
In the QCDf framework in the large recoil limit the seven independent form factors can be expanded in ratios of $1/m_b$ and $1/E_{K^*}$ \cite{Charles98,Beneke00}. 
While $\alpha_s$ corrections to the form factors in QCDf have been 
calculated \cite{Beneke00}, the $1/m_b$ corrections are unknown. 
Neglecting $1/m_b$ corrections, the transversity amplitudes at NLO in $\alpha_s$ in the large recoil limit are\footnote{For $A_t$ and $A_S$, the $\alpha_s$ correction which emerges from describing $A_0$ in terms of $\xi_{\parallel}$ is very small and has been neglected.} \cite{Kruger:2005ep,Bobeth:2008ij,Altmannshofer:2008dz}:
\begin{subequations}\label{eq:TA-NLO}
\begin{align}
A_{\perp}^{L,R}  &=  N \sqrt{2} \sqrt{\lambda} \bigg[ 
\left[ (C_9 + C_9^{\prime}) \mp (C_{10} + C_{10}^{\prime}) \right] \frac{ {V(q^2)} }{ M_B + m_{K^*}} 
+ \frac{2m_b}{q^2} {\cal{T}}_{\perp}^{+}
\bigg]\;, \\
A_{\parallel}^{L,R}  & = - N \sqrt{2}(M_B^2 - m_{K^*}^2) \bigg[ \left[ (C_9 - C_9^{\prime}) \mp (C_{10} - C_{10}^{\prime}) \right] 
\frac{{A_1(q^2)}}{M_B-m_{K^*}}
\nonumber\\
& \qquad +\frac{4 m_b}{M_B} \frac{E_{K^*}}{q^2}{\cal{T}}_{\perp}^{-} 
\bigg]\;,
\end{align}
\begin{align}
A_{0}^{L,R}  &=  - \frac{N}{2 m_{K^*} \sqrt{q^2}}  \bigg \{ 
 \left[ (C_9 - C_9^{\prime}) \mp (C_{10} - C_{10}^{\prime}) \right]
\nonumber\\
 & \qquad \times 
\bigg[ (M_B^2 - m_{K^*}^2 - q^2) ( M_B + m_{K^*}) {A_1(q^2)} 
 -\lambda \frac{{A_2(q^2)}}{M_B + m_{K^*}}
\bigg] 
\nonumber\\
& \qquad + {2 m_b}\bigg[\frac{2 E_{K^*}}{M_B}
 (M_B^2 + 3 m_{K^*}^2 - q^2) {\cal{T}}_{\perp}^{-}
-\frac{\lambda}{M_B^2 - m_{K^*}^2} \bigg( {\cal{T}}_{\perp}^{-}+{\cal{T}}_{\parallel}^{-} \bigg) \bigg]
\bigg\}\;\\
 A_t  &= \frac{N}{\sqrt{q^2}} \sqrt{\lambda} \left[ 2 (C_{10} - C_{10}^{\prime}) + \frac{q^2}{m_{\ell} m_b} (C_{Q_2} - C_{Q_2}^\prime)  \right] \frac{E_{K^*}}{m_{K^*}}{\xi_{\parallel}(q^2)},\\
 A_S  &= - \frac{2N}{m_b} \sqrt{\lambda} (C_{Q_1} - C_{Q_1}^\prime)  \frac{E_{K^*}}{m_{K^*}}{\xi_{\parallel}(q^2)},
\end{align}
\end{subequations}
where $E_{K^*}$ is the energy of the final vector meson in the $B$ rest frame
\begin{align}
E_{K^*} = \frac{M_B^2 +m_{K^*}^2 - q^2}{2 M_B}\;,
\end{align}
and
\begin{align}
  N & = \Bigg[ \frac{G_F^2 \alpha_{em}^2}{3\cdot 2^{10}\, \pi^5 M_B^3} 
     |V_{tb} V_{ts}^\ast|^2\, q^2 \, \sqrt{\lambda} \, \beta_{\ell} \Bigg]^{1/2}\;, &
  \beta_{\ell} & = \sqrt{1-\frac{4 m_{\ell}^2}{q^2}}\;,
\label{eq:betal}
\end{align}
and $\lambda$ is the triangle function $\lambda(M_B^2 , m_{K^*}^2 , q^2)$
\begin{equation}
\lambda= M_B^4  + m_{K^*}^4 + q^4 - 2 (M_B^2 m_{K^*}^2+ m_{K^*}^2 q^2  + M_B^2 q^2)\; .
\label{lambda-kinematic}
\end{equation}
Further explanation on what we use for ${\cal{T}}_{\perp,\parallel}^{\pm}$, $\xi_{\perp,\parallel}$, $A_{1,2}$ and $V$ is given in Appendix \ref{app:b}.

For the differential distribution of the CP conjugate decay $B ^0 \to K ^*(\to K^+ \pi^- ) \ell^+ \ell^-$, if we keep 
the definition of $\theta_{\ell}$ to remain the same and replace $K^-$ with $K^+$
for the definition of $\theta_{K^*}$ and consider $\phi$ to be the angle between the normals of the $\ell^+ \ell^-$ plane and 
the $K^+ \pi^-$ in the $B^0$ rest frame, 
we can use Eq.~(\ref{eq:diffAD}) where $J$ is replaced with $\bar J$. The function $\bar J(q^2, \theta_\ell, \theta_{K^*}, \phi)$
is obtained from Eq.~(\ref{eq:J}) by the replacements \cite{Kruger:1999xa}
\begin{align}
  \label{eq:CP:I}
  J_{1,2}^{(c,s)} & \to  \bar{J}_{1,2}^{(c,s)}, & J_{6}^{(c,s)} & \to -\bar{J}_{6}^{(c,s)} ,\\
  J_{3,4,7} & \to  \bar{J}_{3,4,7}, & J_{5,8,9} & \to -\bar{J}_{5,8,9} ,
\end{align}
where $\bar J$ is equal to $J$ with all the weak phases conjugated. 
The change of sign for $J_{5,6,8,9}$ can be understood by considering that the CP conjugate 
decay leads to the transformations $\theta_l \to \pi -\theta_l$ and $\phi \to -\phi$ in Eq.~(\ref{eq:J}).

\newpage
\subsubsection{Observables}
\label{sec:Observables} 

Integrating Eq. (\ref{eq:diffAD}) over all angles, the dilepton mass distribution can be written in terms of 
the angular coefficients \cite{Bobeth:2008ij,Altmannshofer:2008dz}:
\begin{equation}
\frac{d\Gamma}{dq^2} = \frac{3}{4} \bigg( J_1 - \frac{J_2}{3} \bigg)\;.
\label{eq:dBR}
\end{equation}
For the normalised forward-backward asymmetry $A_{FB}$ we use \cite{Bobeth:2008ij,Altmannshofer:2008dz}:
\begin{equation}
A_{\rm FB}(q^2)  \equiv
     \left[\int_{-1}^0 - \int_{0}^1 \right] d\cos\theta_l\, 
          \frac{d^2\Gamma}{dq^2 \, d\cos\theta_l} \Bigg/\frac{d\Gamma}{dq^2}
          =  -\frac{3}{8} J_6 \Bigg/ \frac{d\Gamma}{dq^2}\; .
\label{eq:AFB}
\end{equation} 
The longitudinal polarisation fraction $F_L$ can be constructed as the 
ratios of the transversity amplitudes and therefore contains less theoretical uncertainty from the form factors.
For the longitudinal polarisation, $F_L$ we have \cite{Kruger:2005ep,Altmannshofer:2008dz}:
\begin{align}
 F_L(s)
&= \frac{-J_2^c}{d\Gamma / dq^2}
\end{align}
where $J_i \equiv 2 J^s_i + J^c_i$ for $i=1,2,6$.

An interesting observable is the zero--crossing of the forward-backward asymmetry ($q_0^2$).
Neglecting the lepton masses and not considering the chirality flipped and scalar operators for the zero of $A_{FB}$ we obtain
\begin{equation}
 q_0^2 = \frac{- m_b}{C_9} \text{Re}\left[ (M_B-m_{K^*})\frac{2 E_{K^*}}{M_B}\frac{{\cal T}_{\perp}}{A_1}+
(M_B+m_{K^*})\frac{{\cal T}_{\perp}^*}{V} \right] .
\label{eq:q0}
\end{equation}
To calculate the zero of $A_{FB}$ in our numerical analysis we have directly used the relation for $A_{FB}$ (Eq.~(\ref{eq:AFB})).

Another observable which is rather independent of hadronic input parameters is the isospin asymmetry which arises from non-factorizable effects.
The non-factorizable effects depend on the charge of the spectator quark, and hence depending on whether the decaying B meson is charged or neutral, there will
be a difference in the contribution of these effects to the decay width which can cause an isospin asymmetry. 
The (CP averaged) isospin asymmetry is defined as \cite{Feldmann:2002iw}
\begin{eqnarray}
\frac{dA_I}{dq^2} &= & 
\frac{ d\Gamma[B^0\to K^{\ast0}\ell^+ \ell^-]/dq^2 -
d\Gamma[B^\pm\to  K^{\ast\pm}\ell^+ \ell^-]/dq^2}
{ d\Gamma[ B^0\to K^{\ast0}\ell^+ \ell^-]/dq^2 +
d\Gamma[B^\pm \to  K^{\ast\pm}\ell^+ \ell^-]/dq^2} \ .
\label{isospin}
\end{eqnarray}
In the SM, $dA_I/dq^2$ is less than 1\% \cite{Feldmann:2002iw,Beneke:2004dp}, 
the smallness of the isospin asymmetry makes it sensitive to isospin-violating NP effects. 

Other observables of interest are the transverse amplitudes which have a small dependence on the form factors and a large 
sensitivity to right-handed currents via $C_7^{\prime}$. They are defined as \cite{Kruger:2005ep,Egede:2008uy,Altmannshofer:2008dz,Matias:2012xw}

\begin{align}
A^{(2)}_{T}({s}) &= \frac{J_3}{2\,J^s_2}\;, \label{eq:AT2}\\
A^{(3)}_{T}({s}) &= \left(\frac{4(J_4)^2+\beta_\ell^2(J_7)^2}{-2 J^c_2\,(2J^s_2+J_3)}\right)^{1/2}\;,\label{eq:AT3}\\
A^{(4)}_{T}({s}) &= \left(\frac{\beta_\ell^2(J_5)^2 + 4(J_8)^2}{4 (J_4)^2 + \beta_\ell^2 (J_7)^2}\right)^{1/2}\;,\label{eq:AT4}
\end{align}
as well as
\begin{align}
A_{Im}({s}) &= \frac{J_9}{d\Gamma / dq^2}\;,
\end{align}
which is sensitive to the complex phases, but very small ($\mathcal{O}(10^{-3})$) in the SM.

All the observables can also be expressed in terms of the CP averaged angular coefficients $S_i$ \cite{Altmannshofer:2008dz}. In particular, $S_3$, which has recently been measured by LHCb, can be related to $F_L$ and $A^{(2)}_{T}$ by
\begin{align}
 S_3 = \frac12 (1 - F_L) A^{(2)}_{T}\;.
\end{align} 

For the high $q^2$ region we follow \cite{Bobeth:2010wg}, where the heavy quark effective theory (HQET) of \cite{Grinstein:2004vb} is used.
In that framework improved Isgur-Wise form factor relations are used to extrapolate the lacking high $q^2$ form factors from the calculations of the low
$q^2$ form factors and an operator product expansion in powers of $1/Q^2$, ($Q=(m_b,\sqrt{q^2})$), is used for the estimation of the long-distance effects
from quark loops. For the low recoil region (high $q^2$), the following observables are proposed by \cite{Bobeth:2010wg} 
\begin{align}
  H_T^{(1)} & = \frac{\sqrt{2} J_4}{\sqrt{- J_2^c \left(2 J_2^s - J_3\right)}} \;,
  \label{eq:def:HT1}
\\
  H_T^{(2)} & = \frac{\beta_l J_5}{\sqrt{-2 J_2^c \left(2 J_2^s + J_3\right)}} \;,
  \label{eq:def:HT2}
\\
  H_T^{(3)} & = \frac{\beta_l J_6}{2 \sqrt{(2 J_2^s)^2 - (J_3)^2}}\;.
  \label{eq:HT}
\end{align} 
These transversity observables have small hadronic uncertainties at high $q^2$.

\subsubsection{Standard Model values and theoretical uncertainties}

For our numerical analysis we have used the NLO relations for the transversity amplitudes in the ($1<q^2<6$ GeV$^2$) region where we are 
sufficiently below the charm resonance threshold ($q^2=m_c^2$) and far enough from the kinematic minimum where the decay amplitude is dominated by the photon pole.
The input parameters can be found in Table~\ref{tab:input} and the SM Wilson coefficients are given in Table~\ref{tab:wilson}.
\begin{table}
\begin{center}
\footnotesize{\begin{tabular}{|l|l|l|l|l|l|l|l|l|l|}\hline
$C_1(\mu_b)$ & $C_2(\mu_b)$ & $C_3(\mu_b)$ & $C_4(\mu_b)$ & $C_5(\mu_b)$ & $C_6(\mu_b)$ & $C_7^{\text{eff}}(\mu_b)$ & $C_8^{\text{eff}}(\mu_b)$ & $C_9(\mu_b)$ & $C_{10}(\mu_b)$  \\ \hline
$-0.2610$    & $1.0076$     & $-0.0052$    & $-0.0795$    & $0.0004$     & $0.0009$     & $-0.2974$                 & $-0.1614$                 & $4.2297$     & $-4.2068$  \\ \hline
\end{tabular}}
\caption{SM Wilson coefficients at $\mu_b=m_b^{\text{pole}}$ and $\mu_0=2M_W$ to NNLO accuracy in $\alpha_s$. \label{tab:wilson}}
\end{center}
\end{table}
The available experimental values are given for $q^2$ bins which can be shown as $\langle \text{observable} \rangle_{[q^2_{min},q^2_{max}]}$, 
in other words the $dq^2$ integration is over the $[q^2_{min},q^2_{max}]$ bin.
For normalised quantities 
like $A_{FB}$ and $F_L$ the numerator and denominator are integrated separately \cite{Bobeth:2010wg,Kruger:2005ep}:
\begin{align}
\langle A_{FB} \rangle_{[q^2_{min},q^2_{max}]} &= -\frac{3}{8} \frac{\langle J_6 \rangle_{[q^2_{min},q^2_{max}]}}{\langle {d\Gamma}/{dq^2} \rangle_{[q^2_{min},q^2_{max}]}} ,\\
\langle F_L \rangle_{[q^2_{min},q^2_{max}]} &=  \frac{\langle -J_2^c \rangle_{[q^2_{min},q^2_{max}]}}{\langle {d\Gamma}/{dq^2} \rangle_{[q^2_{min},q^2_{max}]}}. 
\end{align}

\begin{table}
\begin{center}
\begin{tabular}{|l|l|l|l|l|l|l|}\hline 
  Observable                                                                & SM prediction & (FF)       & (SL)        & (QM)                & (CKM) & (Scale) \\ \hline \hline
  $10^7 \times  BR(B \to K^* \mu^+ \mu^-)_{[1,6]}$ & $2.32$        & $\pm 1.34$ & $\pm 0.04$  & $ _{-0.03}^{+0.04}$ & $_{-0.13}^{+0.08}$ & $_{-0.05}^{+0.09}$  \\ \hline
  $\langle A_{FB}(B \to K^* \mu^+ \mu^-) \rangle_{[1,6]}$         & $-0.06$        & $\pm 0.04$ & $\pm 0.02$  & $ \pm 0.01$ & --- & ---  \\ \hline
  $\langle F_{L}(B \to K^* \mu^+ \mu^-) \rangle_{[1,6]}$          & $0.71$        & $\pm 0.13$ & $\pm 0.01$  & $ \pm 0.01 $ & --- & ---\\ \hline
  $q_0^2 (B \to K^* \mu^+ \mu^-)/\text{GeV}^2 $                                & $4.26$        & $\pm 0.30$ & $\pm 0.15$  & $_{-0.04}^{+0.14}$  & --- & $_{-0.04}^{+0.02}$\\ \hline
 \end{tabular}
\caption{SM predictions and errors. \label{tab:errors}}
\end{center}
\end{table}

We estimate the theoretical uncertainties for the SM values in the low $q^2$ region using two methods: one follows the approach of \cite{Bobeth:2011}, and the second is based on a Monte Carlo method. Both methods give similar results. We consider five different sources of uncertainty.
The errors from the form factors (FF) have been calculated by adding in quadrature the uncertainties due to $V$, $A_1$ and $A_2$ 
(11\%, 13\% and 14\%, respectively \cite{Ball:2004rg}).
For the unknown $1/m_b$ sub-leading corrections (SL), we have assumed 7\% corrections 
to the $\cal{T}_{\perp,\parallel}$ terms in the transversity amplitudes\footnote{If for the (SL) error a 10\% correction to the $\cal{T}_{\perp,\parallel}$ is considered the overall uncertainties will not have a significant change since they are dominated by the (FF) uncertainties.}, 
these corrections have been added in quadrature.
Another group of errors is from the quark mass uncertainties (QM)
which we have estimated by
separately varying $m_t^{pole}$, $m_b^{\overline{MS}}$ and $m_c^{\overline{MS}}$ according to Table~\ref{tab:input} and 
 added in quadrature. Another source of error comes from the uncertainty in the CKM matrix element 
combination $|V_{tb}V_{ts}^*|$ (CKM) which has been computed by considering the uncertainty given in Table~\ref{tab:input}.
 The last source of error that we consider is the scale dependence (Scale) which we estimate by varying 
$\mu_b$ between ${\mu_b}/2$ and $2{\mu_b}$ (with ${\mu_b}=m_b^{pole}$).
  These five groups of errors for
$BR$, $A_{FB}$, $F_L$ and $q_0^2$ in the SM have been gathered in Table~\ref{tab:errors}.
For the Standard Model predictions the primed coefficients ($C_{i,Q_1,Q_2}^{\prime}$) as well as the scalar coefficients ($C_{Q_{1,2}}$) have been put to zero.

For the high $q^2$ region we have used the relative errors of Table 2 in \cite{Bobeth:2010wg}.

\section{Experimental results}

At present, the best upper limit for $\mathrm{BR}(B_s\to\mu^+\mu^-)$ measured in a single experiment comes from LHCb \cite{Aaij:2012ac}:
\begin{equation}
\mathrm{BR}(B_s\to\mu^+\mu^-) < 4.5 \times 10^{-9}
\label{bsmumu}
\end{equation}
at 95\% C.L. This upper limit is followed by the result from CMS \cite{Chatrchyan:2012rg}:
\begin{equation}
\mathrm{BR}(B_s\to\mu^+\mu^-) < 7.7 \times 10^{-9}\;.
\end{equation}
The CDF collaboration obtains a 95\% C.L. upper limit \cite{Aaltonen:2011fi}: 
\begin{equation}
\mathrm{BR}(B_s\to\mu^+\mu^-) < 3.4 \times 10^{-8}\;,
\end{equation}
together with a one sigma interval 
\begin{equation}
\mathrm{BR}(B_s\to\mu^+\mu^-)  = (1.3^{+0.9}_{-0.7})\times10^{-8}\;,
\end{equation}
coming from an observed excess over the expected background.

For our numerical evaluations, we consider the LHCb limit, and
accounting for 11\% theoretical uncertainty (as explained in
section~\ref{sec:Bsmumu}), we impose the following limit at 95\%
  C.L. 
\begin{equation}
\mathrm{BR}(B_s\to\mu^+\mu^-) < 5.0 \times 10^{-9}\;.
\end{equation}

For $B\to K^* \mu^+\mu^-$ related observables we also use the latest LHCb results
which correspond to an integrated luminosity of 1 fb$^{-1}$ \cite{LHCb-CONF-2012-008}.
The results are summarised in Table \ref{tab:experiment}
where the experimental uncertainties are statistical and systematic. For comparison, the SM predictions with the corresponding theoretical errors (from the five sources of errors
mentioned in Table \ref{tab:errors}, added in quadrature) are also provided.

\begin{table}
\begin{center}
\begin{tabular}{|l|l|l|l|l|l|l|}\hline 
  Observable                                                                & SM prediction & Experiment       \\ \hline \hline
  $10^7 \text{GeV}^2 \times \langle dBR/dq^2\; (B \to K^* \mu^+ \mu^-) \rangle_{[1,6]}$ & $0.47 \pm 0.27 $        & $0.42 \pm 0.04 \pm 0.04$   \\ \hline
  $\langle A_{FB}(B \to K^* \mu^+ \mu^-) \rangle_{[1,6]}$         & $-0.06 \pm 0.05 $        & $-0.18 ^{+0.06+0.01}_{-0.06-0.01}$   \\ \hline
  $\langle F_{L}(B \to K^* \mu^+ \mu^-) \rangle_{[1,6]}$          & $0.71 \pm 0.13 $        & $0.66 ^{+0.06+0.04}_{-0.06-0.03}$   \\ \hline
  $q_0^2 (B \to K^* \mu^+ \mu^-)/\text{GeV}^2$      & $4.26 ^{+0.36}_{-0.34} $        & $4.9  ^{+1.1}_{-1.3}$   \\ \hline
 \end{tabular}
\caption{Experimental values and SM predictions (the theoretical errors are from adding in quadrature the 
different errors in Table \ref{tab:errors}). \label{tab:experiment}}
\end{center}
\end{table}

In addition to the observables in Table \ref{tab:experiment}, three other observables
have also been measured using 1 fb$^{-1}$ of LHCb data namely $S_3$
and $A_{Im}$ \cite{LHCb-CONF-2012-008} and also very recently the isospin asymmetry $A_{I}$ \cite{isospin}. The reported results are:
\begin{eqnarray}
\langle 2\,S_3(B \to K^* \mu^+ \mu^-) \rangle_{[1,6]} &=& 0.10^{+0.15+0.02}_{-0.16-0.01}\;,\\
\langle A_{Im}(B \to K^* \mu^+ \mu^-) \rangle_{[1,6]} &=& 0.07^{+0.07+0.02}_{-0.7-0.01}\;,\\
\langle A_{I}(B \to K^* \mu^+ \mu^-) \rangle_{[1,6]} &=& -0.15 \pm 0.16\;.\label{eq:AI}
\end{eqnarray}

However, with the current experimental accuracy, these observables are not sensitive enough to probe SUSY parameters. 

\section{Constraints on SUSY}
To illustrate the impact of the recent limit on BR($B_s\to \mu^+\mu^-$) and the measurements of the angular distributions and differential branching fraction of the $B\to K^*\mu^+\mu^-$ decay by LHCb, we consider constrained MSSM scenarios. The observables are calculated as described in section~\ref{sec:2} using SuperIso v3.3 \cite{Mahmoudi:2007vz,Mahmoudi:2008tp}. 

We focus on two specific scenarios, both assuming SUSY breaking mediated by gravity and invoking unification boundary conditions at a very high scale $m_\mathrm{GUT}$ where the universal mass parameters are specified. The first model is the CMSSM, characterised by the set of parameters $\{m_0,m_{1/2},A_0,\tan\beta,\mathrm{sgn}(\mu)\}$. Here $m_0$ is the universal mass of the scalars, $m_{1/2}$ the universal gaugino mass, $A_0$ the universal trilinear coupling, and $\tan\beta$ the ratio of the vacuum expectation values of the Higgs doublets. The second model we consider involves non-universal Higgs masses (NUHM). This model generalises the CMSSM, allowing for the GUT scale mass parameters of the Higgs doublets to have values different from $m_0$. The two additional parameters can be traded for two other parameters at a lower scale, conveniently the $\mu$ parameter and the mass $M_A$ of the $CP$-odd Higgs boson implying that the charged Higgs boson mass can be treated essentially as a free parameter, contrary to the CMSSM.

\begin{figure}[t!]
\begin{center}
\includegraphics[width=7.cm]{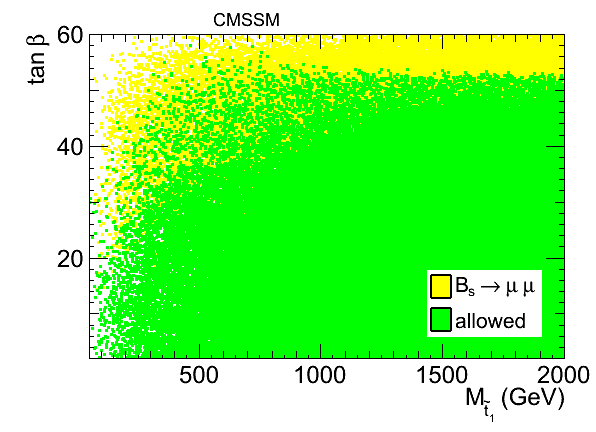}\quad\includegraphics[width=7.cm]{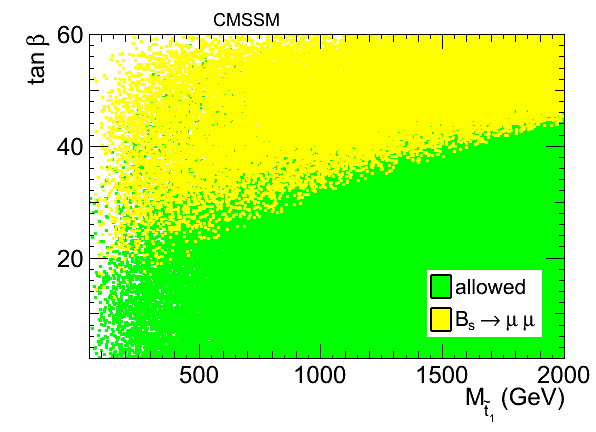}\\
\includegraphics[width=7.cm]{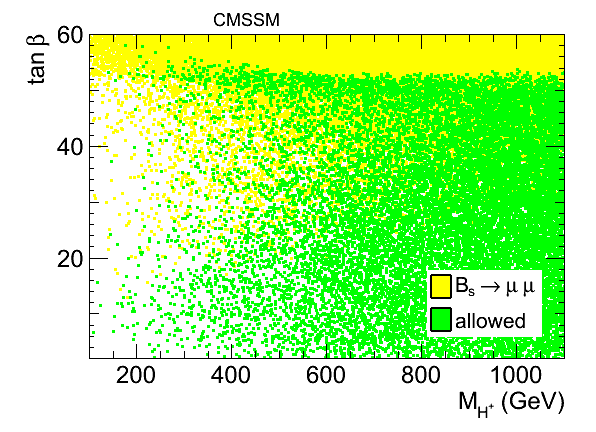}\quad\includegraphics[width=7.cm]{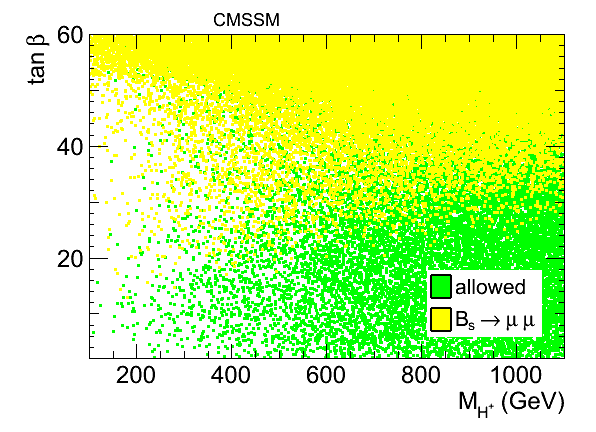}
\caption{Constraint from BR($B_s\to \mu^+\mu^-$) in the CMSSM plane ($M_{\tilde t_1}$, $\tan\beta$) in the upper panel and ($M_{H^{\pm}}$, $\tan\beta$) in the lower panel, with the allowed points displayed in the foreground in the left and in the background in the right. \label{fig:Bsmumu-CMSSM}}
\end{center}
\end{figure}

First we study the constraints from BR($B_s\to \mu^+\mu^-$) in the CMSSM. For this purpose, we scanned over  
$m_0 \in [100,3000]$ GeV, $m_{1/2} \in [100,3000]$ GeV, $A_0 \in [-5000,2000]$ GeV and $\tan\beta \in [1,60]$, with $\mu >0$. We generated about 500,000 points, and for each point we calculate the spectrum of SUSY particle masses and couplings using SOFTSUSY 3.2.4 \cite{Allanach:2001kg} and compute BR($B_s\to \mu^+\mu^-$) using SuperIso v3.3. The constraints are shown in Fig.~\ref{fig:Bsmumu-CMSSM} in the planes ($M_{\tilde t_1}$, $\tan\beta$) and ($M_{H^{\pm}}$, $\tan\beta$). These results could be compared to the previous results using the LHCb 2010 collected data \cite{Akeroyd:2011kd,LHCb:2011ac}. It is clear that the large values of $\tan\beta$ are strongly constrained.

\begin{figure}[t!]
\begin{center}
\includegraphics[width=7.cm]{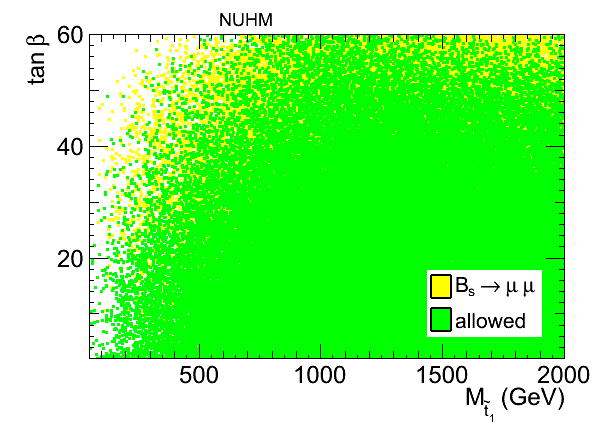}\quad\includegraphics[width=7.cm]{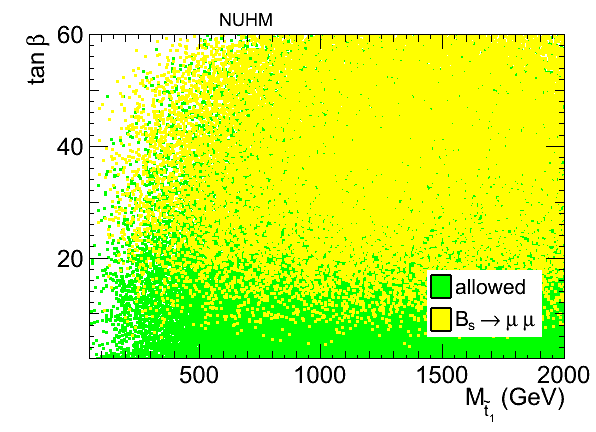}\\
\includegraphics[width=7.cm]{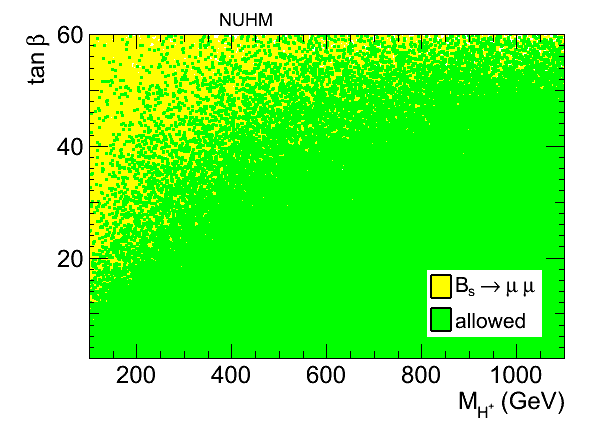}\quad\includegraphics[width=7.cm]{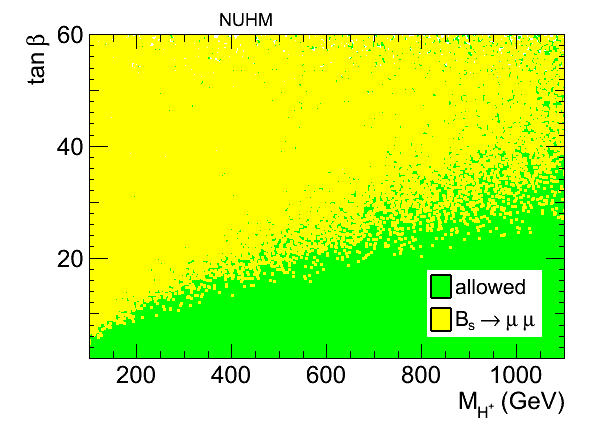}
\caption{Constraint from BR($B_s\to \mu^+\mu^-$) in the NUHM plane ($M_{\tilde t_1}$, $\tan\beta$) in the upper panel and ($M_{H^{\pm}}$, $\tan\beta$) in the lower panel, with the allowed points displayed in the foreground in the left and in the background in the right. \label{fig:Bsmumu-NUHM}}
\end{center}
\end{figure}

In the NUHM scenario, the parameters are scanned over in a similar way as in the CMSSM, and in addition $\mu \in [-2000,2000]$ GeV and $M_A \in [50,1100]$ GeV. The results are shown in Fig.~\ref{fig:Bsmumu-NUHM}. Since there are two additional degrees of freedom in NUHM as compared to CMSSM, it is easier for a model point to escape the constraint, as can be seen by comparing Figs.~\ref{fig:Bsmumu-CMSSM} and \ref{fig:Bsmumu-NUHM} where the allowed points are displayed on top. On the other hand since $M_{\tilde t_1}$ and $M_{H^{\pm}}$ are discorrelated in NUHM, it is possible for any $M_{\tilde t_1}$ to have a very small $M_{H^{\pm}}$ and therefore being excluded, as can be seen from the plot with the allowed points in the background.

The effect of an SM--like measurement of BR($B_s\to \mu^+\mu^-$) in non--constrained MSSM (the pMSSM) is demonstrated in \cite{Arbey:2011aa}.
 
\begin{figure}[!t]
\begin{center}
\includegraphics[width=6.1cm]{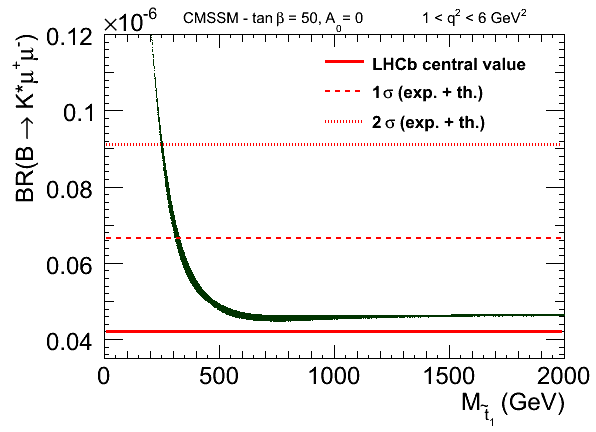}\quad\includegraphics[width=6.1cm]{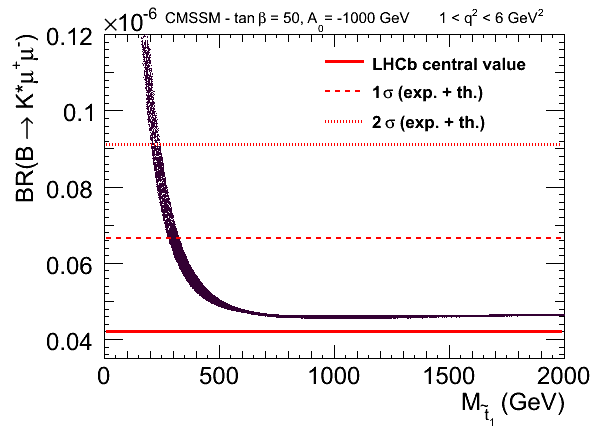}\\
\includegraphics[width=6.1cm]{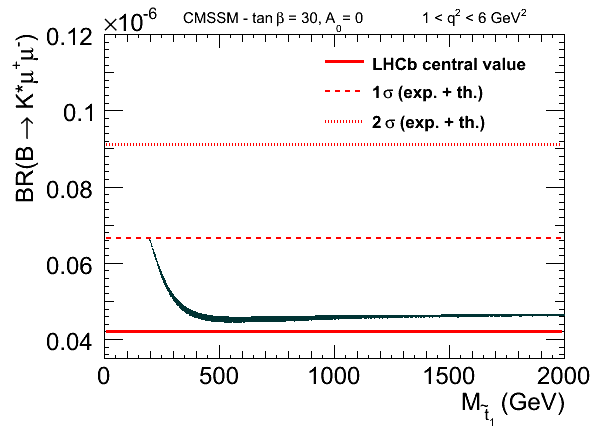}\quad\includegraphics[width=6.1cm]{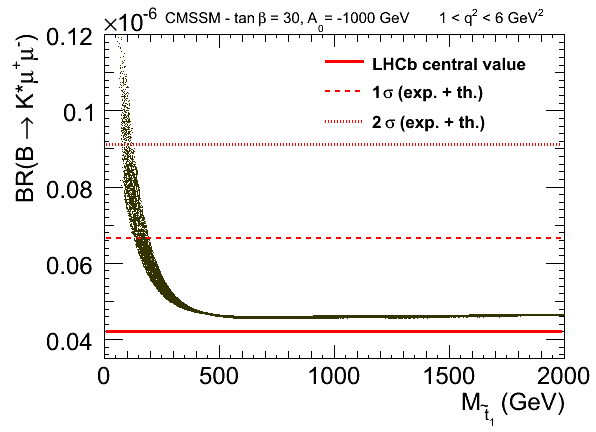}
\vspace*{-0.2cm}
\caption{SUSY spread of the averaged BR($B\to K^*\mu^+\mu^-$) at low $q^2$ in function of the lightest stop mass, for $\tan\beta$=50 (upper panel) and $\tan\beta$=30 (lower panel), in the left for $A_0=0$ and in the right for $A_0=-1000$ GeV. \label{fig:BRlow}}
\vspace*{0.4cm}
\includegraphics[width=6.1cm]{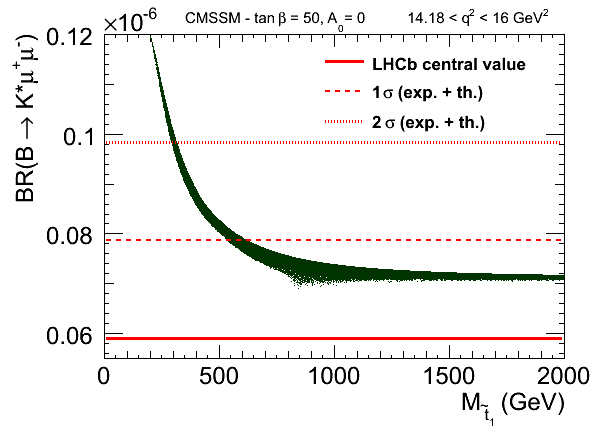}\quad\includegraphics[width=6.1cm]{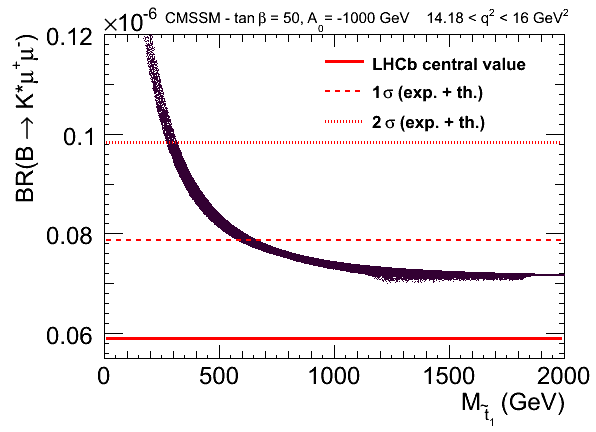}\\
\includegraphics[width=6.1cm]{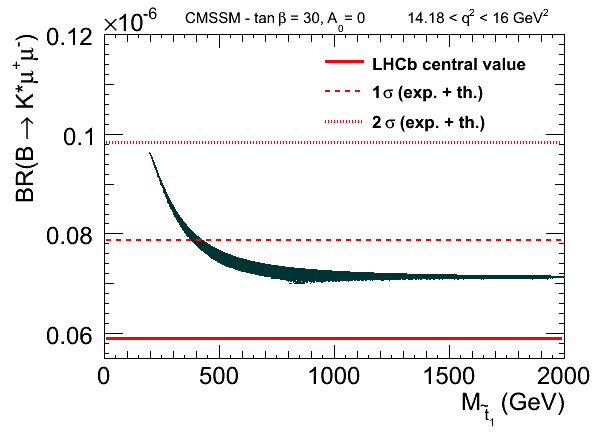}\quad\includegraphics[width=6.1cm]{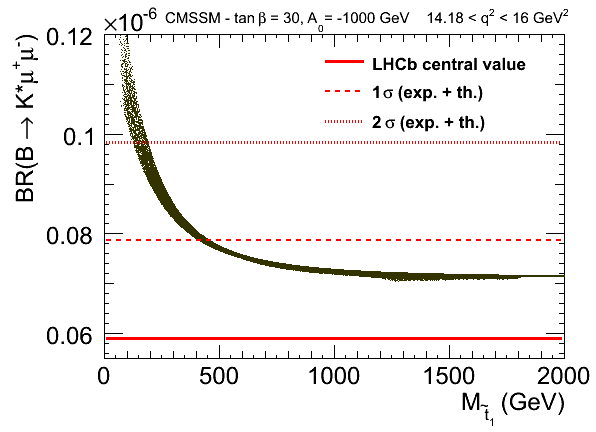}
\vspace*{-0.2cm}
\caption{SUSY spread of the averaged BR($B\to K^*\mu^+\mu^-$) at high $q^2$, similar to Fig.~\ref{fig:BRlow}. \label{fig:BRhigh}}
\end{center}
\end{figure}
\begin{figure}[!t]
\begin{center}
\includegraphics[width=6.1cm]{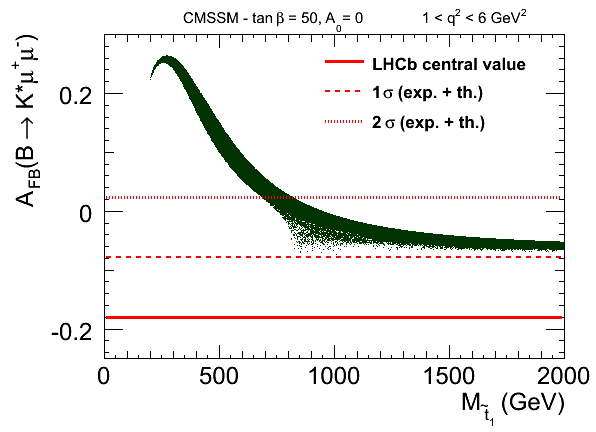}\quad\includegraphics[width=6.1cm]{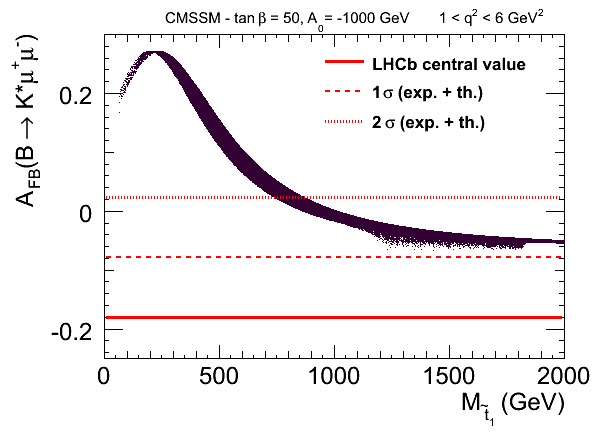}\\
\includegraphics[width=6.1cm]{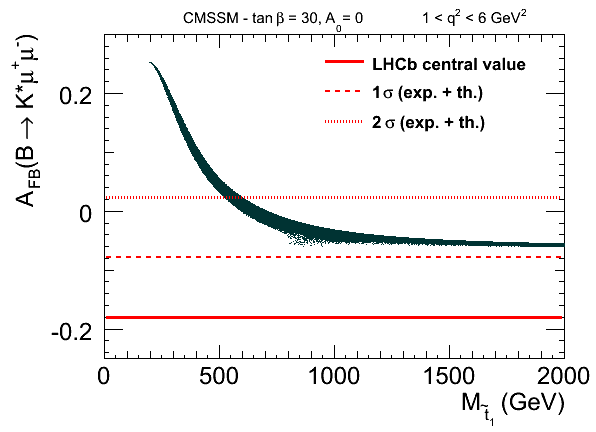}\quad\includegraphics[width=6.1cm]{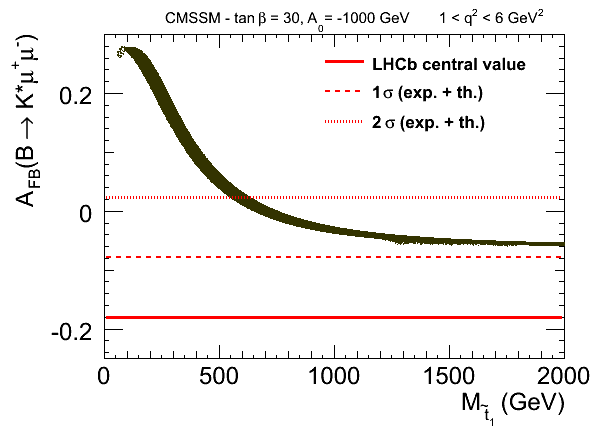}
\vspace*{-0.2cm}
\caption{SUSY spread of the $A_{FB}(B\to K^*\mu^+\mu^-)$ at low $q^2$ in function of the lightest stop mass, for $\tan\beta$=50 (upper panel) and $\tan\beta$=30 (lower panel), in the left for $A_0=0$ and in the right for $A_0=-1000$ GeV. \label{fig:AFBlow}}
\vspace*{0.4cm}
\includegraphics[width=6.1cm]{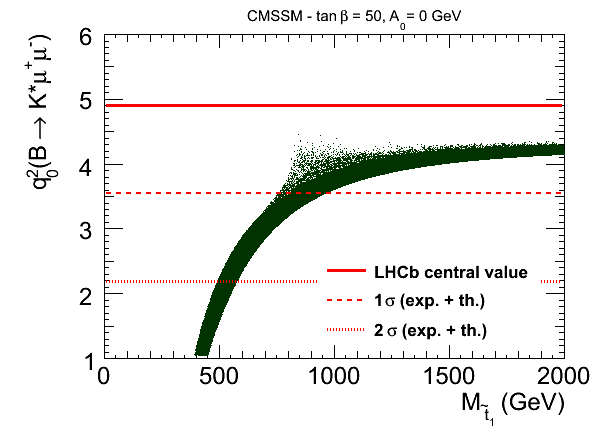}\quad\includegraphics[width=6.1cm]{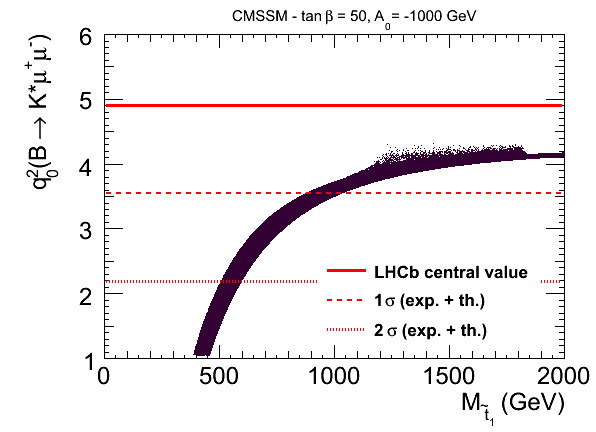}\\
\includegraphics[width=6.1cm]{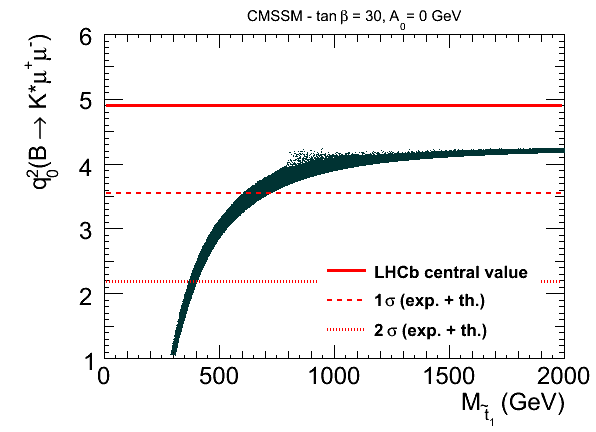}\quad\includegraphics[width=6.1cm]{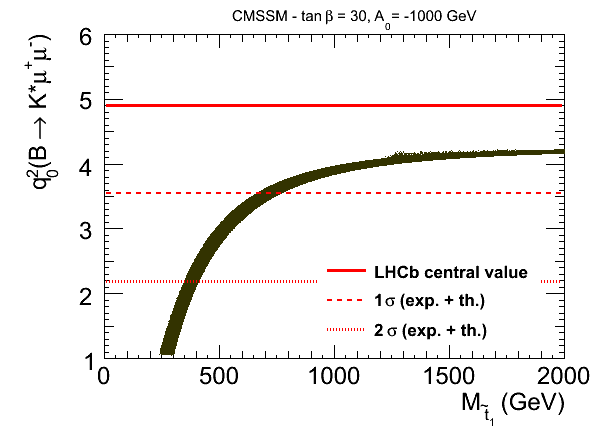}
\vspace*{-0.2cm}
\caption{SUSY spread of the $A_{FB}(B\to K^*\mu^+\mu^-)$ zero--crossing, similar to Fig.~\ref{fig:AFBlow}. \label{fig:AFB0}}
\end{center}
\end{figure}
\begin{figure}[t!]
\begin{center}
\includegraphics[width=6.1cm]{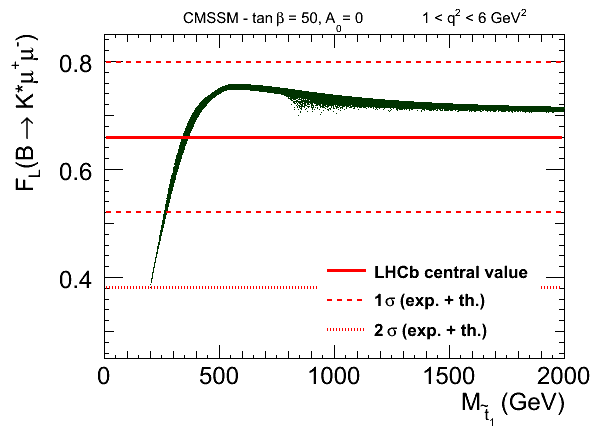}\quad\includegraphics[width=6.1cm]{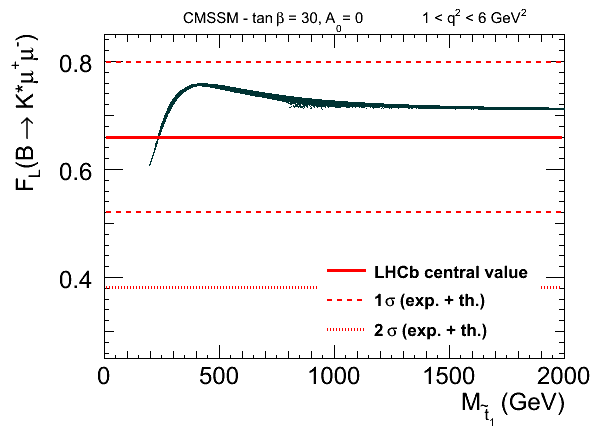}\\
\includegraphics[width=6.1cm]{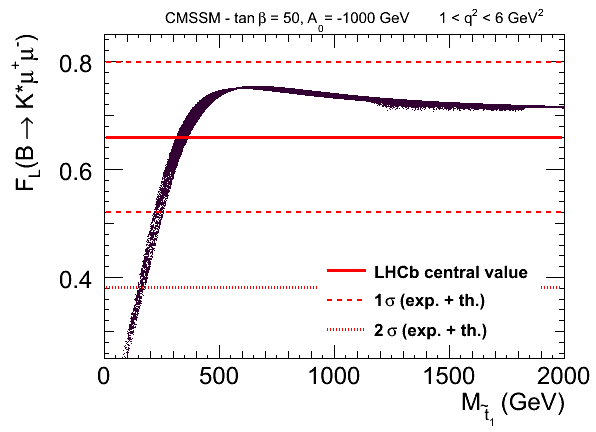}\quad\includegraphics[width=6.1cm]{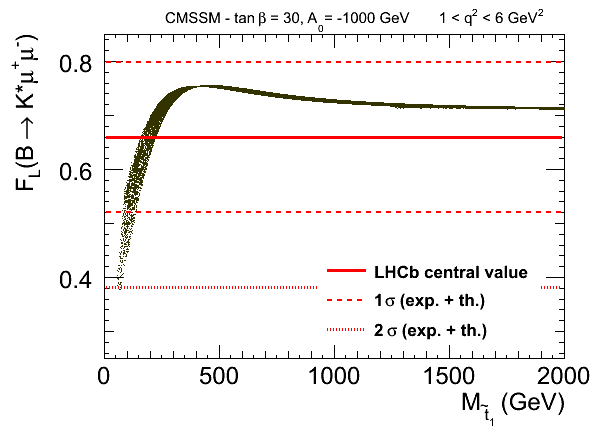}
\caption{SUSY spread of the $F_L(B\to K^*\mu^+\mu^-)$ at low $q^2$ in function of the lightest stop mass, for $\tan\beta$=50 (upper panel) and $\tan\beta$=30 (lower panel), in the left for $A_0=0$ and in the right for $A_0=-1000$ GeV. \label{fig:FLlow}}
\end{center}
\end{figure}
%
Next we consider the constraints from $B\to K^*\mu^+\mu^-$ observables.
In order to study the maximal effects we consider $\tan\beta$=50 but show also the results for $\tan\beta$=30, and investigate the SUSY spread in function of the lightest stop mass. We start with the averaged differential branching ratio as defined in Table~\ref{tab:errors}. The results in CMSSM are displayed in Fig.~\ref{fig:BRlow} for the low $q^2$ region and in Fig.~\ref{fig:BRhigh} for the high $q^2$ region, where the solid red lines correspond to the LHCb central value, while the dashed and dotted lines represent the 1 and 2$\sigma$ bounds respectively, including both theoretical and experimental errors (added in quadrature). At low $q^2$, this branching ratio excludes $M_{\tilde t_1}$ below $\sim$ 250 GeV for $\tan\beta$=50 and $\sim$ 150 GeV for $\tan\beta$=30. In the high $q^2$ region the branching ratio is doing slightly better, as the $M_{\tilde t_1}$ below $\sim$ 300 GeV and $\sim$ 200 GeV are excluded for $\tan\beta$=50 and $\tan\beta$=30 respectively.
As this light stop region is already excluded by the direct SUSY searches for the same scenario, BR($B\to K^*\mu^+\mu^-$) does not provide additional information. The main reason of the limited constraining power of the branching ratio is the large theoretical uncertainties (mainly due to form factors) from which this observable is suffering. 
The results are shown for two values of $A_0$ (=0 and -1000 GeV) for comparison. As can be seen from the figures, the constraints with smaller $A_0$ are slightly stronger.

Contrary to the branching ratio, angular distributions, in which the theoretical uncertainties are reduced, can in principle provide more robust constraints on the SUSY parameter space. In particular, we consider in the following the forward-backward asymmetry $A_{FB}$, the zero-crossing $q^2_0$ of $A_{FB}$, $F_L$, as well as $S_3$, $A_{Im}$ for which the LHCb results with 1~fb$^{-1}$ of data are available. The two latter observables do not provide any constraint with the current results and accuracy. The SUSY spread in function of the stop mass of $A_{FB}$, $q^2_0$ and $F_L$ is given in Figs.~\ref{fig:AFBlow}--\ref{fig:FLlow}. As can be seen, $A_{FB}$ provides the most stringent constraints among these observables, and excludes $M_{\tilde t_1} \lesssim$ 800 GeV and 600 GeV at $\tan\beta$=50 and $\tan\beta$=30 respectively. $q^2_0$ on the other hand excludes $M_{\tilde t_1} \lesssim$ 550 GeV (for $\tan\beta$=50) and 400 GeV (for $\tan\beta$=30) while $F_L$ excludes $M_{\tilde t_1} \lesssim$ 200 GeV (for $\tan\beta$=50) and 150 GeV (for $\tan\beta$=30). The impressive constraining power of $A_{FB}$ is mainly due to the fact that the measured central value is smaller than the SM prediction and in addition the reported experimental errors are more than twice smaller than the previous results.

Same observables at high $q^2$ have less impact on the SUSY parameters and therefore their results are not reproduced here.

\begin{figure}[t!]
\begin{center}
\includegraphics[width=6.1cm]{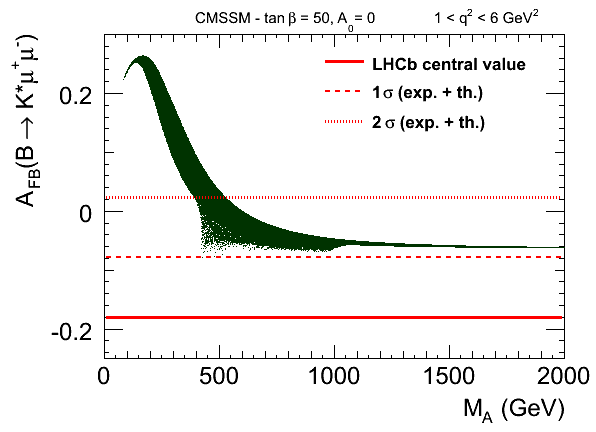}\quad\includegraphics[width=6.1cm]{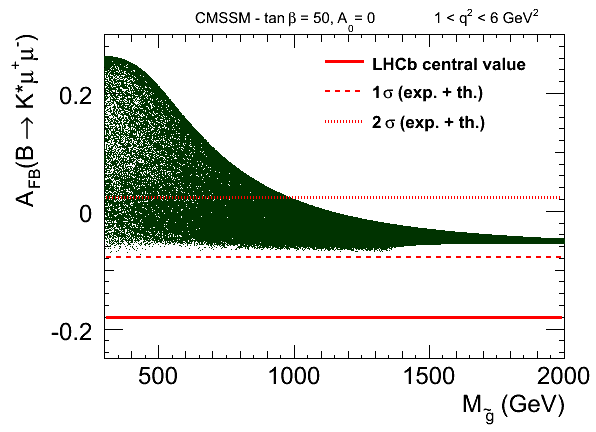}
\caption{SUSY spread of the $A_{FB}(B\to K^*\mu^+\mu^-)$  at low $q^2$ in function of the pseudo scalar Higgs mass (left) and the gluino mass (right), for $\tan\beta$=50 and $A_0=0$. \label{fig:AFB_MA_Mgl}}
\end{center}
\end{figure}

\newpage
To illustrate the sensitivity of the observables to other SUSY parameters, in Fig.~\ref{fig:AFB_MA_Mgl} we present the variation of $A_{FB}$ (which has the largest impact as we have shown) with respect to the pseudo scalar Higgs mass and the gluino mass.

\begin{figure}[t!]
\begin{center}
\includegraphics[width=6.1cm]{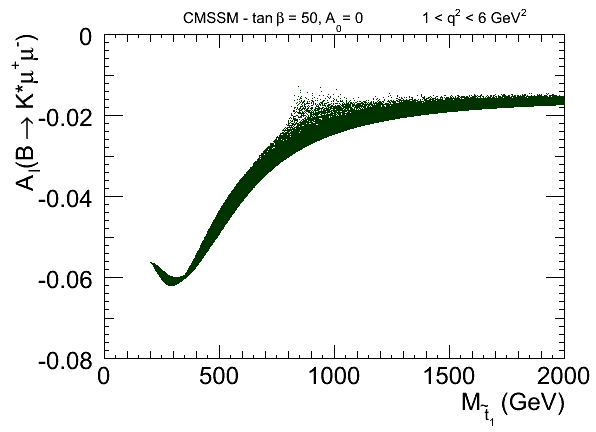}
\caption{SUSY spread of the isospin asymmetry $(B\to K^*\mu^+\mu^-)$ at low $q^2$ in function of the lightest stop mass, for $\tan\beta$=50 and $A_0=0$. \label{fig:AI_low}}
\end{center}
\end{figure}

\newpage
Another observable of interest for which LHCb has announced the measurement very recently \cite{isospin} is the isospin asymmetry $A_I$, defined in Eq.~(\ref{isospin}). Fig.~\ref{fig:AI_low} illustrates the SUSY spread of $A_I$ in function of $M_{\tilde t_1}$. Since the experimental measurement (Eq.~(\ref{eq:AI})) has an error larger than the SUSY spread, the current result does not provide any information on the SUSY parameters. 

\begin{figure}[!t]
\begin{center}
\includegraphics[width=5.5cm]{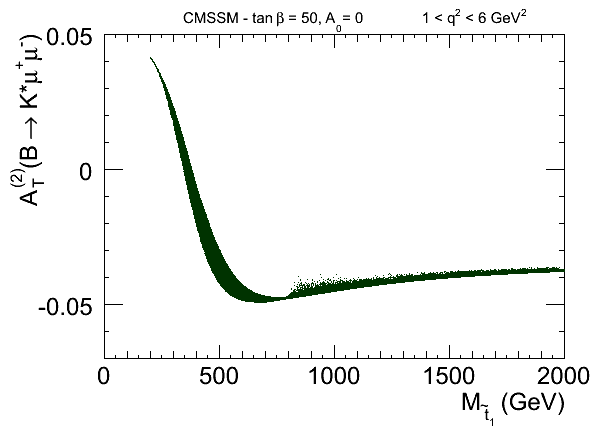}
\includegraphics[width=5.5cm]{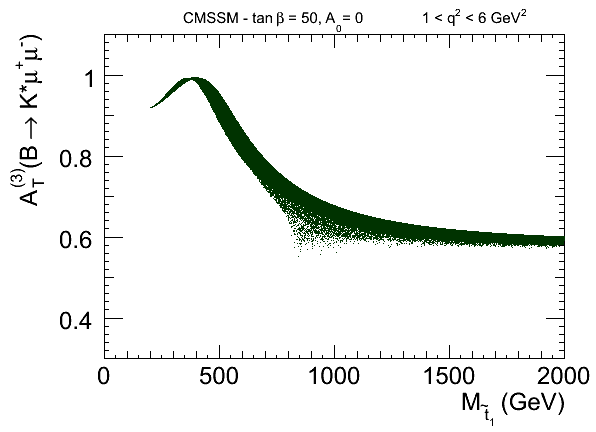}
\includegraphics[width=5.5cm]{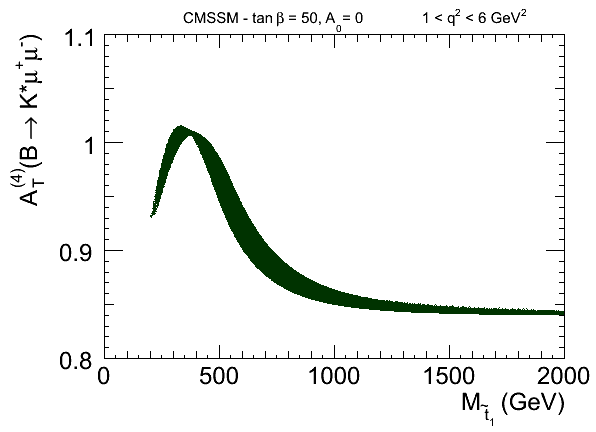}
\caption{SUSY spread of $A_T^{(2)}$, $A_T^{(3)}$ and $A_T^{(4)}$ at low $q^2$ in function of the lightest stop mass, for $\tan\beta$=50 and $A_0=0$.}
\label{fig:AT}
\end{center}
\end{figure}

In Fig.~\ref{fig:AT}, we show the SUSY spread in function of the lightest stop mass, for $A_T^{(2)}$, $A_T^{(3)}$, $A_T^{(4)}$ (defined in Eqs.~(\ref{eq:AT2})--(\ref{eq:AT4})) which are not measured yet by LHCb. Large deviations from the Standard Model can be seen for small values of $M_{\tilde t_1}$, and depending on their eventual upcoming experimental measurements they could be of interest in constraining SUSY parameters. 

\begin{figure}[!t]
\begin{center}
\includegraphics[width=7.cm]{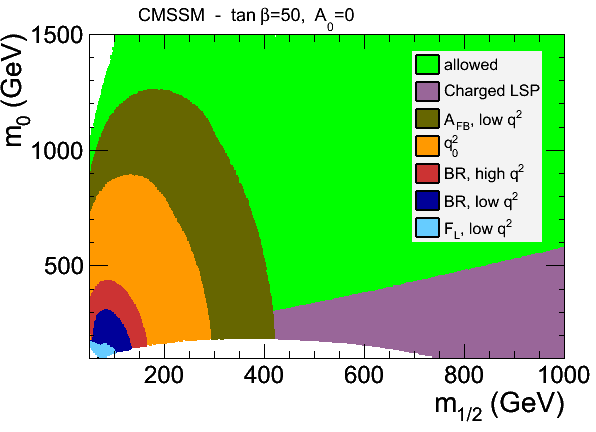}\quad\includegraphics[width=7.cm]{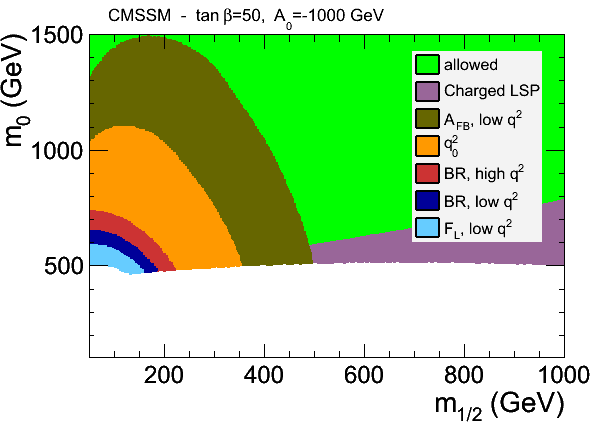}\\
\includegraphics[width=7.cm]{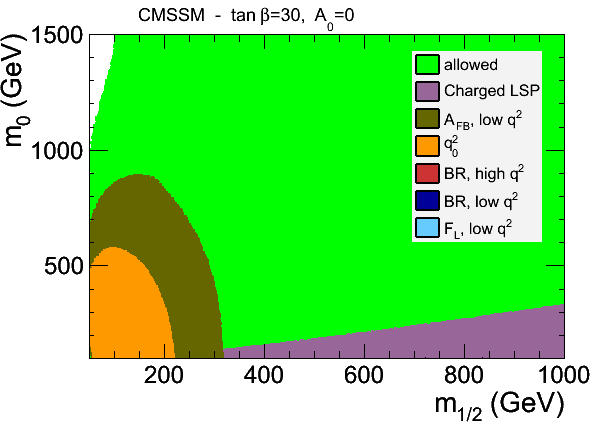}\quad\includegraphics[width=7.cm]{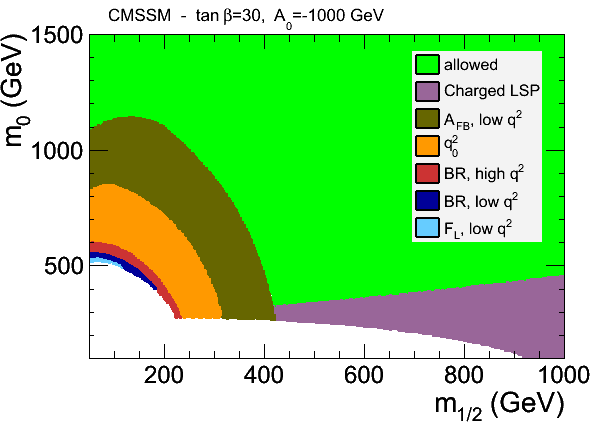}
\caption{Constraining power of the $B\to K^*\mu^+\mu^-$ observables in the plane ($m_{1/2},m_0$), for $\tan\beta$=50 (upper panel) and $\tan\beta$=30 (lower panel), in the left for $A_0=0$ and in the right for $A_0=-1000$ GeV.}
\label{fig:m0m12}
\end{center}
\end{figure}

A comparison between these observables in the plane ($m_{1/2},m_0$) is given in Fig.~\ref{fig:m0m12}. As expected, $A_{FB}$ is the most constraining observable also in this plane. All the observables show more sensitivity at larger $\tan\beta$ and smaller $A_0$.

\begin{figure}[!t]
\begin{center}
\includegraphics[width=7.cm]{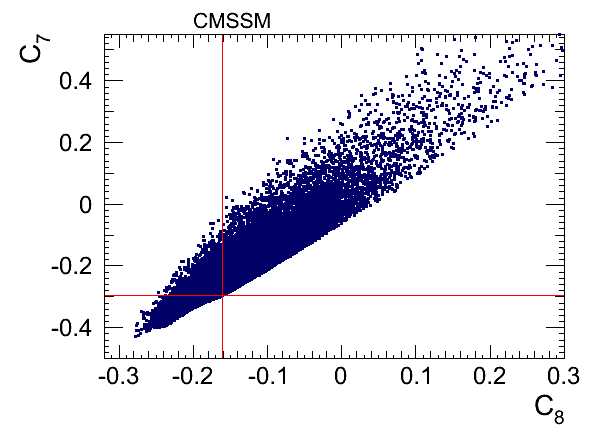}\quad\includegraphics[width=7.cm]{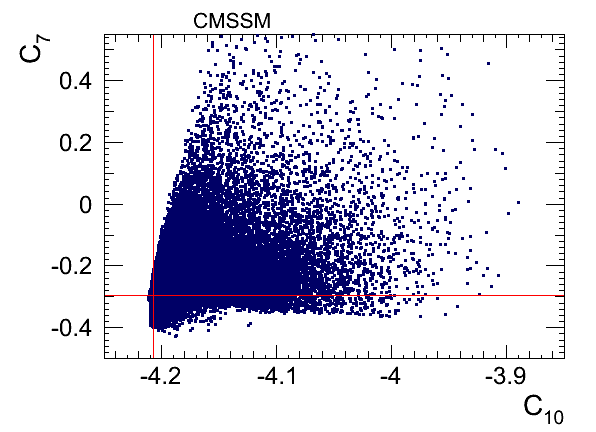}\\
\includegraphics[width=7.cm]{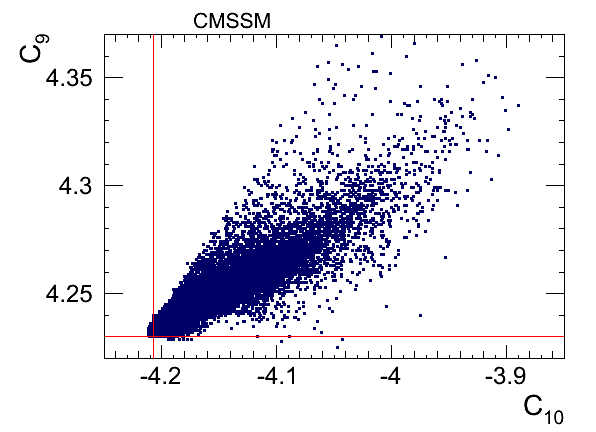}\quad\includegraphics[width=7.cm]{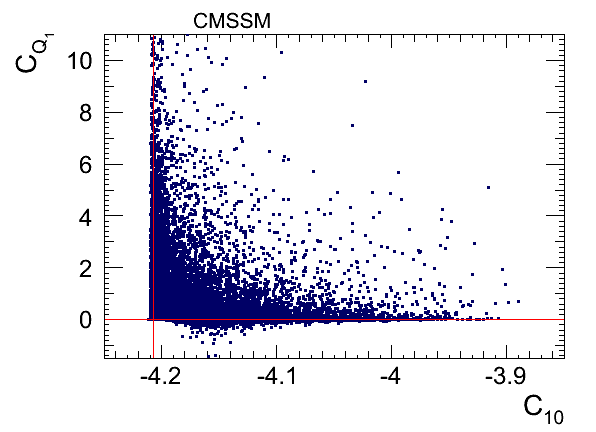}
\caption{Variation of the Wilson coefficients in the CMSSM with all the parameters varied in the ranges given in the text. The red lines correspond to the SM predictions.}
\label{fig:WC}
\end{center}
\end{figure}

\begin{figure}[!t]
\begin{center}
\includegraphics[width=7.cm]{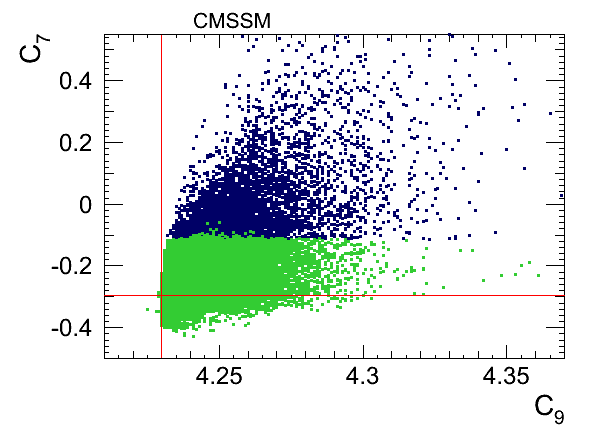}
\caption{Variation of $C_7$ and $C_9$ in the CMSSM with all the parameters varied in the ranges given in the text. Only the green points are allowed by the zero--crossing of $A_{FB}$. The red lines correspond to the SM predictions.}
\label{fig:C7q0}
\end{center}
\end{figure}

To trace the origin of these constraints, it is interesting to present them in terms of Wilson coefficients. The variation of the observables in SUSY is driven on the one hand by the additional contributions to the Wilson Coefficients, in particular $C_7$, $C_8$, $C_9$ and $C_{10}$, and on the other hand by new contributions through the scalar and pseudo scalar coefficients $C_{Q_{1}}$ and $C_{Q_{2}}$. The full variation of the Wilson coefficients in a full CMSSM scan is presented in Fig.~\ref{fig:WC}, ignoring existing constraints on SUSY parameters or Wilson coefficients. As can be seen, $C_7$ and $C_8$ can have both signs in SUSY and their correlation is visible in the figure. As for $C_9$, it varies only by very little while $C_{10}$ can have a larger spread. This feature can be understood once we notice that box diagrams are suppressed with respect to scalar- or $Z$-penguin diagrams, giving $\delta C_9/\delta C_{10}\sim (g_V/g_A)_\mu\sim1-4\sin^2\theta_W$. On the other hand, scalar coefficients can receive very large contributions in SUSY, as already known.
An interesting information can be obtained from the zero--crossing of $A_{FB}$ since it depends on the relative sign of $C_7$ and $C_9$ as can be seen in Eq.~(\ref{eq:q0}), and excludes the positive sign of $C_7$ as can be seen in Fig.~\ref{fig:C7q0}.

\begin{figure}[!t]
\begin{center}
\includegraphics[width=7.cm]{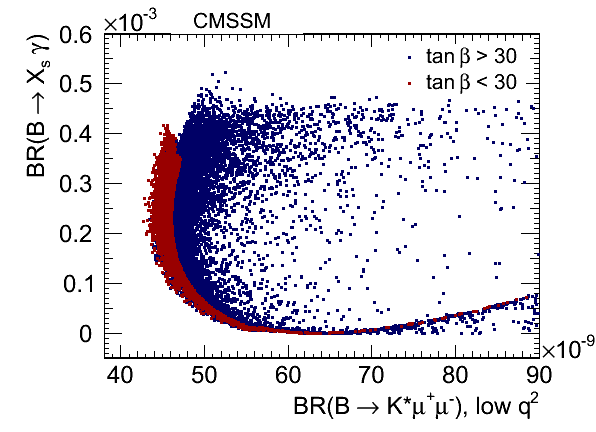}\quad\includegraphics[width=7.cm]{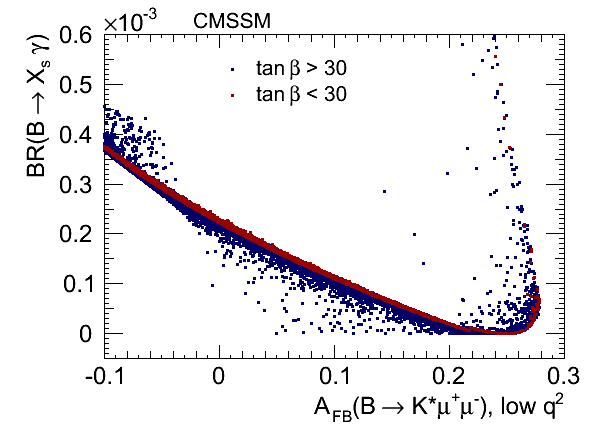}
\caption{The correlation between BR($B\to X_s\gamma$) and averaged BR($B\to K^*\mu^+\mu^-$) in the left, and between  BR($B\to X_s\gamma$) and $A_{FB}(B\to K^*\mu^+\mu^-$) in the right.}
\label{fig:BRs}
\end{center}
\end{figure}

Finally, in Fig.~\ref{fig:BRs} we show the correlations of BR($B\to X_s\gamma$) with BR($B\to K^*\mu^+\mu^-$) and $A_{FB}(B\to K^*\mu^+\mu^-)$. In the SM, BR($B\to X_s\gamma$) is dominated by contributions from $C_7$. The SM prediction for this branching ratio is $(3.08 \pm 0.22) \times 10^{-4}$ \cite{Misiak:2006zs,Misiak:2006ab,Mahmoudi:2008tp} while the latest combined experimental value from HFAG is $(3.55 \pm 0.24 \pm 0.09) \times 10^{-4}$ \cite{Asner:2010qj}.
As expected, there are strong correlations between BR($B\to X_s\gamma$) and $B\to K^*\mu^+\mu^-$ branching ratio and forward backward asymmetry, especially for small and intermediate values of $\tan \beta$, where the scalar and pseudo scalar contributions have a limited effect.

\newpage~\newpage
\section{Conclusions}
The rare decays $B_s\to\mu^+\mu^-$ and $B\to K^*\mu^+\mu^-$ are of great importance in exploring the footprints of physics beyond the SM. We have presented in this paper an update of the constraints by the recent LHCb limit on BR($B_s\to\mu^+\mu^-$) in two constrained MSSM scenarios, CMSSM and NUHM. 

We also demonstrated that $B\to K^*\mu^+\mu^-$ provides a large number of complementary observables from the full angular distribution. In particular while $B_s\to\mu^+\mu^-$ lose very rapidly its sensitivity by lowering down $\tan\beta$, the constraints by $B\to K^*\mu^+\mu^-$ observables weaken only mildly and thus act as additional tools to explore SUSY parameters. The forward backward asymmetry is in particular a very powerful observable which provides competitive information to the direct SUSY searches. To display the origin of these constraints in the CMSSM, we have shown the possible values and correlations between the Wilson coefficients of the various operators relevant for these processes. Among those, the Wilson coefficient $C_9$ is a parameter for which these angular observables provide a unique handle. However, we have shown that the sensitivity of $C_9$ to CMSSM effects is suppressed by the muon vector coupling, because they proceed mainly through a $Z$ penguin diagram. The constraining power of other observables, like the isospin asymmetry whose measurement is under way, have been shown to display their interest in the framework of Minimal Flavour Violation models for which the CMSSM provides an interesting prototype.

 With reduced theoretical and experimental errors for these observables, in particular for the forward backward asymmetry and its zero--crossing which have been recently measured for the first time by the LHCb experiment, one would have access to very powerful observables in constraining supersymmetric models. The information obtained in this way can also serve as consistency checks with the results of the direct searches.

\begin{appendix}

\section{Angular coefficients $J_i$}
\label{app:a}
The angular coefficients are given below \cite{Altmannshofer:2008dz}.
\begin{subequations}
\begin{align}
\label{eq:J_i}
  J_1^s & = \frac{(2+\beta_\ell^2)}{4} \left[|{A_\perp^L}|^2 + |{A_\parallel^L}|^2 + (L\to R) \right] 
            + \frac{4 m_\ell^2}{q^2} \text{Re}\left({A_\perp^L}^{}{A_\perp^R}^* + {A_\parallel^L}^{}{A_\parallel^R}^*\right)\;, 
\\
  J_1^c & =  |{A_0^L}|^2 +|A_0^R|^2  + \frac{4m_\ell^2}{q^2} 
               \left[|A_t|^2 + 2\text{Re}({A_0^L}^{}{A_0^R}^*) \right] + \beta_\ell^2 |A_S|^2 \;,
\\
  J_2^s & = \frac{ \beta_\ell^2}{4}\left[ |{A_\perp^L}|^2+ |{A_\parallel^L}|^2 + (L\to R)\right]\;,
\\
  J_2^c & = - \beta_\ell^2\left[|{A_0^L}|^2 + (L\to R)\right]\;,
\\
  J_3 & = \frac{1}{2}\beta_\ell^2\left[ |{A_\perp^L}|^2 - |{A_\parallel^L}|^2  + (L\to R)\right]\;,
\\
  J_4 & = \frac{1}{\sqrt{2}}\beta_\ell^2\left[\text{Re} ({A_0^L}^{}{A_\parallel^L}^*) + (L\to R)\right]\;,
\\
  J_5 & = \sqrt{2}\beta_\ell\left[\text{Re}({A_0^L}^{}{A_\perp^L}^*) - (L\to R) 
- \frac{m_\ell}{\sqrt{q^2}}\, \text{Re}({A_\parallel^L} {A_S^*}+{A_\parallel^R} {A_S^*})
\right]\;,
\\
  J_6^s  & = 2\beta_\ell\left[\text{Re} ({A_\parallel^L}^{}{A_\perp^L}^*) - (L\to R) \right]\;,
\\
   J_6^c  &  =
 4 \beta_\ell  \frac{m_\ell}{\sqrt{q^2}}\, \text{Re} \left[ {A_0^L} {A_S^*} + (L\to R) \right]\;,
\\
  J_7 & = \sqrt{2} \beta_\ell \left[\text{Im} ({A_0^L}^{}{A_\parallel^L}^*) - (L\to R) 
+ \frac{m_\ell}{\sqrt{q^2}}\, {\text{Im}}({A_\perp^L} {A_S^*}+{A_\perp^R} {A_S^*})
\right]\;,
\\
  J_8 & = \frac{1}{\sqrt{2}}\beta_\ell^2\left[\text{Im}({A_0^L}^{}{A_\perp^L}^*) + (L\to R)\right],
\\
  J_9 & = \beta_\ell^2\left[\text{Im} ({A_\parallel^L}^{*}{A_\perp^L}) + (L\to R)\right]\;.
\end{align}
\end{subequations}
with $\beta_{\ell}$ defined in Eq.~(\ref{eq:betal}).

\section{Form factors}
\label{app:b}
The ${\cal T}_{\perp,\parallel}$ (which include non-factorisable corrections) are described in terms of the soft form factors and can be found in
\cite{Beneke:2001at,Beneke:2004dp} where, in order to get ${\cal{T}}_{\perp,\parallel}^{\pm}$, $C_7^{\text{eff}}$ appearing in ${\cal T}_{\perp,\parallel}$ 
should be replaced with $(C_7^{\text{eff}}\pm C_7^{\prime})$
\cite{Bobeth:2008ij}.
To obtain the soft form factors we have used the factorisation scheme used in \cite{Beneke:2004dp}:
\begin{align}
  \xi_\perp (q^2) & = \frac{M_B}{M_B + m_{K^*}} V(q^2)\;,\\
  \xi_\parallel (q^2) & = \frac{M_B + m_{K^*}}{2 E_{K^*}} A_1 (q^2) -\frac{M_B - m_{K^*}}{M_B} A_2 (q^2)\;.
\end{align}
The full form factors $V$ and $A_{1,2}$ have been taken from light-cone sum rule (LCSR) calculations \cite{Ball:2004rg}:
\begin{align}
V(q^2) &= \frac{r_1}{1 - q^2/m_R^2} + \frac{r_2}{1 - q^2/m_{fit}^2} \;,\\ 
A_1(q^2) &= \frac{r_2}{1 - q^2/m_{fit}^2} \;,\\ 
A_2(q^2) &= \frac{r_1}{1 - q^2/m_{fit}^2} + \frac{r_2}{(1 - q^2/m_{fit}^2)^2} \;,
\label{eq:qdependent-ff}
\end{align}
where the fit parameters $r_{1,2}, m^2_{R}$ and $m^2_{fit}$ are given in Table \ref{tab:FF:fit}.
\begin{table}[t]
\centering
\begin{tabular}{|c|cccc|}
\hline
  $ $ & $r_1$ & $r_2$ & $m_R^2\,[\text{GeV}^2]$ & $m_{fit}^2\,[\text{GeV}^2]$ \\
\hline
  $V$   & $0.923$  & $-0.511$ & $5.32^2$ & $49.40$ \\[0.5ex]
  $A_1$ &  \mbox{} & $ 0.290$ &  \mbox{} & $40.38$ \\[0.5ex]
  $A_2$ & $-0.084$ &  $0.342$ &  \mbox{} & $52.00$ \\[0.5ex]
\hline
\end{tabular}
\caption{\label{tab:FF:fit} Fit parameters describing the $q^2$ dependence
 of the form factors $V$ and $A_{1,2}$ in the LCSR approach \cite{Ball:2004rg}.}
\end{table}

\section{Renormalisation group equations for $C_9$}
\label{app:c}
To obtain the evolved Wilson coefficients from the matching scale ($\mu_0$) to $\mu_b$, the renormalisation group equation (RGE) 
for $\vec{\widetilde{C}}$ has to be solved:
\begin{align}
 \mu \frac{d}{d \mu} \vec{\widetilde{C}}(\mu) = \hat{\gamma}^T(g) \vec{\widetilde{C}}(\mu) \;,
\end{align}
where $\hat{\gamma}$ is the Anomalous Dimension Matrix (ADM). A general solution for the RGE is
\begin{align}
 \vec{\widetilde{C}}(\mu_b) = \hat{U}(\mu_b,\mu_0) \vec{\widetilde{C}}(\mu_0)  \;, 
\end{align}
where $\hat{U}$ is the evolution matrix
\begin{align}
 \hat{U}(\mu_b,\mu_0) = T_g\exp \int_{g(\mu_0)}^{g(\mu_b)} dg' \frac{(\hat{\gamma}^T(g'))}{\beta(g')} \;.
\end{align}

$T_g$ is the g($=\sqrt{4\pi {\alpha}_s}$)-ordering operator such that the coupling constants increase from right to left. $\hat{\gamma}$ 
and $\beta$ can be expanded in g 

\begin{align}\label{eq:}
 \hat{\gamma}(g)&=\frac{g^2}{16\pi^2} \hat{\gamma }^{(0)} + \left(\frac{g^2}{16\pi^2}\right)^2 \hat{\gamma }^{(1)} + \left(\frac{g^2}{16\pi^2}\right)^3 \hat{\gamma }^{(2)} + \text{...}\\
\beta(g)&=-\frac{g^3}{16\pi^2}\beta_0 -\frac{g^5}{\left(16\pi^2\right)^2}\beta_1 -\frac{g^7}{\left(16\pi^2\right)^4}\beta _2 +\text{...}
\end{align}

The Wilson coefficients and the evolution matrix expanded in $\alpha_s$ are
\begin{align}
 \hat{U}(\mu_b,\mu_0) &= \sum_{n\geqslant 0} \left(\frac{\alpha_s(\mu_0)}{4\pi}\right)^{(n)} \hat{U}^{(n)}\\
 \vec{\widetilde{C}}_i(\mu) &=  \sum_{n\geqslant 0} \left(\frac{\alpha_s(\mu)}{4\pi}\right)^{(n)} \vec{\widetilde{C}}^{(n)}_i(\mu).
\end{align}

Following the conventions used in \cite{Czakon:2006ss}, the evolution matrix is
\begin{align} \label{eq:Ukl}
 U^{(n)}_{kl} = \sum_{j=0}^n \sum_{i=1}^{9} m^{(nj)}_{kli} \eta^{a_i-j}\\
\end{align}
where $\eta=\alpha_s(\mu_0)/\alpha_s(\mu_b)$.

Using the above relations for the different orders of $C_9(\mu_b)$ we have
\begin{align}
\tilde{C}^{(0)}_9(\mu_b) &= \sum_{l=1}^{9} U^{(0)}_{9l} \tilde{C}^{(0)}_l(\mu_0)\\
\tilde{C}^{(1)}_9(\mu_b) &= \eta \left[ 
\sum_{l=1}^{9} U^{(0)}_{9l} \tilde{C}^{(1)}_l(\mu_0)+
\sum_{l=1}^{9} U^{(1)}_{9l} \tilde{C}^{(0)}_l(\mu_0)\right]\\
\tilde{C}^{(2)}_9(\mu_b) &= \eta^2 \left[ 
\sum_{l=1}^{9} U^{(0)}_{9l} \tilde{C}^{(2)}_l(\mu_0)+
\sum_{l=1}^{9} U^{(1)}_{9l} \tilde{C}^{(1)}_l(\mu_0)+
\sum_{l=1}^{9} U^{(2)}_{9l} \tilde{C}^{(0)}_l(\mu_0)\right].
\end{align}

To obtain the evolution matrix we have followed \cite{Buras:2011we} and \cite{Buras:2006gb}. 
Taking $\hat{\gamma}^{(0)}$, $\hat{\gamma}^{(1)}$ and $\hat{\gamma}^{(2)}$ from \cite{Czakon:2006ss} and \cite{Gambino:2003zm}, 
we have produced the necessary ``magic numbers'' ($m_{kli}$) for the evaluation of $U_{9l}$.
Since $O_9$ does not mix with $O_{7,8}$ the magic numbers $m_{97i}$ and $m_{98i}$ are all zero.
The $a_i$'s and the rest of the $m_{kli}$ can be found in Tables~\ref{tab:ai}-\ref{tab:magic22}. 
 
In the above formulas, $\tilde{C}^{(n)}_{l}(\mu_0)$ for $l\neq7,8$ can all be found in section 2 of \cite{Bobeth:1999mk} 
where to get our $\tilde{C}^{(n)}_{l}(\mu_0)$ 
we use $\tilde{C}^{(n)}_{l}(\mu_0)={C}^{t(n)}_{l}(\mu_0)-{C}^{c(n)}_{l}(\mu_0)$,
and $\tilde{C}^{(n)}_{7,8}(\mu_0)$ are not needed since $U_{97}$ and $U_{98}$ are zero.
To get the Wilson coefficients based on our choice of operators (Eq.~(\ref{physical_basis})) we have
\begin{align}
 C_9 (\mu_b) &=\frac{4\pi}{\alpha_s(\mu_b)}\tilde{C}_9 .
\end{align}

\begin{table}
\begin{center}
\begin{tabular}{|r|rrrrrrrrr|}\hline
$i$&1&2&3&4&5&6&7&8&9\\\hline
$a_i$ & $\frac{14}{23}$ & $\frac{16}{23}$ & $\frac{6}{23}$ & $-\frac{12}{23}$ & $  0.4086 $ & $ -0.4230 $ & $ -0.8994 $ & $  0.1456 $ & $ -1 $\\[1mm]
\hline
\end{tabular}
\caption{The numbers $a_i$ used in Eq.~(\ref{eq:Ukl}) \label{tab:ai}}
\end{center}
\end{table}

\begin{table}
\begin{center}
\begin{tabular}{|r|rrrrrrrrr|}\hline
$i$&1&2&3&4&5&6&7&8&9\\\hline
$m^{(00)}_{91i}$&0&0&-0.0328&-0.0404&0.0021&-0.0289&-0.0174&-0.0010&0.1183  \\[1mm]
$m^{(00)}_{92i}$&0&0&-0.0985&0.0606&0.0108&0.0346&0.0412&-0.0018&-0.0469 \\[1mm]
$m^{(00)}_{93i}$&0&0&0&0&0.0476&-0.1167&-0.3320&-0.0718&0.4729  \\[1mm]
$m^{(00)}_{94i}$&0&0&0&0&0.0318&0.0918&-0.2700&0.0059&0.1405  \\[1mm]
$m^{(00)}_{95i}$&0&0&0&0&0.9223&-2.4126&-1.5972&-0.4870&3.57455  \\[1mm]
$m^{(00)}_{96i}$&0&0&0&0&0.4245&1.1742&-0.0507&-0.0293&-1.5186  \\[1mm]
$m^{(00)}_{99i}$&0&0&0&0&0&0&0&0&1\\\hline
\end{tabular}
\caption{``Magic numbers'' $m^{(00)}_{9li}.$\label{tab:magic00}}
\end{center}
\end{table}

\begin{table}
\begin{center}
\begin{tabular}{|r|rrrrrrrrr|}\hline
$i$&1&2&3&4&5&6&7&8&9\\\hline
 $m^{(10)}_{91i}$& 0 & 0 & 0.1958 & -0.0442 & -0.0112 & -0.1111 & 0.1283 & 0.0114 & -0.3596 \\[1mm]
 $m^{(10)}_{92i}$& 0 & 0 & 0.2917 & 0.2482 & 0.0382 & 0.1331 & -0.2751 & 0.0260 & -0.8794 \\[1mm]
 $m^{(10)}_{93i}$& 0 & 0 & 0 & 0 & -0.1041 & -0.5696 & 9.5004 & 0.0396 & -0.4856 \\[1mm]
 $m^{(10)}_{94i}$& 0 & 0 & 0 & 0 & -0.0126 & -0.4049 & -0.6870 & 0.1382 & 0.4172  \\[1mm]
 $m^{(10)}_{95i}$& 0 & 0 & 0 & 0 & 4.7639 & -35.0057 & 30.7862 & 5.5105 & 62.3651 \\[1mm]
 $m^{(10)}_{96i}$& 0 & 0 & 0 & 0 & -1.9027 & -1.8789 & -43.9516 & 1.9612 & 54.4557 \\[1mm]
 $m^{(10)}_{99i}$& 0 & 0 & 0 & 0 & 0 & 0 & 0 & 0 & 0  \\
\hline
\end{tabular}
\caption{``Magic numbers'' $m^{(10)}_{9li}.$\label{tab:magic10}}
\end{center}
\end{table}

\begin{table}
\begin{center}
\begin{tabular}{|r|rrrrrrrrr|}\hline
$i$&1&2&3&4&5&6&7&8&9\\\hline
 $m^{(11)}_{91i}$ & 0 & 0 & 0.2918 & 0.0484 & -0.0331 & -0.0269 & 0.0200 & -0.1094 & 0 \\[1mm]
 $m^{(11)}_{92i}$ & 0 & 0 & 0.8754 & -0.0725 & -0.1685 & 0.0323 & -0.0475 & -0.2018 & 0 \\[1mm]
 $m^{(11)}_{93i}$ & 0 & 0 & 0 & 0 & -0.7405 & -0.1088 & 0.3825 & -7.9139 & 0 \\[1mm]
 $m^{(11)}_{94i}$ & 0 & 0 & 0 & 0 & -0.4942 & 0.0856 & 0.3111 & 0.6465 & 0 \\[1mm]
 $m^{(11)}_{95i}$ & 0 & 0 & 0 & 0 & -14.3464 & -2.2495 & 1.8402 & -53.6643 & 0 \\[1mm]
 $m^{(11)}_{96i}$ & 0 & 0 & 0 & 0 & -6.6029 & 1.0948 & 0.0584 & -3.2339 & 0 \\[1mm]
 $m^{(11)}_{99i}$ & 0 & 0 & 0 & 0 & 0 & 0 & 0 & 0 & 0 \\[1mm]
\hline
\end{tabular}
\caption{``Magic numbers'' $m^{(11)}_{9li}.$\label{tab:magic11}}
\end{center}
\end{table}

\begin{table}
\begin{center}
\begin{tabular}{|r|rrrrrrrrr|}\hline
 $i$&1&2&3&4&5&6&7&8&9\\\hline
 $m^{(20)}_{91i}$ & 0 & 0 & 0.6878 & -0.9481 & -0.1928 & -0.8077 & -0.2554 & 0.0562 & -0.6436 \\[1mm]
 $m^{(20)}_{92i}$ & 0 & 0 & 1.3210 & 3.1616 & -0.4814 & 1.9362 & -5.0873 & 0.0468 & -13.5825 \\[1mm]
 $m^{(20)}_{93i}$ & 0 & 0 & 0 & 0 & -2.5758 & -5.8751 & 0.0922 & 0.6433 & 7.7756 \\[1mm]
 $m^{(20)}_{94i}$ & 0 & 0 & 0 & 0 & -2.6194 & 1.1302 & -27.7073 & -0.8550 & 16.0333 \\[1mm]
 $m^{(20)}_{95i}$ & 0 & 0 & 0 & 0 & -6.4519 & -555.931 & 35.1531 & 80.2925 & 102.043 \\[1mm]
 $m^{(20)}_{96i}$ & 0 & 0 & 0 & 0 & -53.3822 & 34.3969 & -124.609 & -32.7515 & -98.8845 \\[1mm]
 $m^{(20)}_{99i}$ & 0 & 0 & 0 & 0 & 0 & 0 & 0 & 0 & 0 \\[1mm]
\hline
\end{tabular}
\caption{``Magic numbers'' $m^{(20)}_{9li}.$\label{tab:magic20}}
\end{center}
\end{table}

\begin{table}
\begin{center}
\begin{tabular}{|r|rrrrrrrrr|}\hline
 $i$&1&2&3&4&5&6&7&8&9\\\hline
 $m^{(21)}_{91i}$ & 0 & 0 & -1.7394 & 0.0530 & 0.1741 & -0.1036 & -0.1478 & 1.2522 & 0 \\
 $m^{(21)}_{92i}$ & 0 & 0 & -2.5918 & -0.2971 & -0.5949 & 0.1241 & 0.3170 & 2.8655 & 0 \\[1mm]
 $m^{(21)}_{93i}$ & 0 & 0 & 0 & 0 & 1.6188 & -0.5311 & -10.9454 & 4.36311 & 0 \\[1mm]
 $m^{(21)}_{94i}$ & 0 & 0 & 0 & 0 & 0.1967 & -0.3775 & 0.7915 & 15.2328 & 0 \\[1mm]
 $m^{(21)}_{95i}$ & 0 & 0 & 0 & 0 & -74.1049 & -32.6399 & -35.4688 & 607.188 & 0 \\[1mm]
 $m^{(21)}_{96i}$ & 0 & 0 & 0 & 0 & 29.5971 & -1.7519 & 50.6366 & 216.094 & 0 \\[1mm]
 $m^{(21)}_{99i}$ & 0 & 0 & 0 & 0 & 0 & 0 & 0 & 0 & 0 \\[1mm]
\hline
\end{tabular}
\caption{``Magic numbers'' $m^{(21)}_{9li}.$\label{tab:magic21}}
\end{center}
\end{table}

\begin{table}
\begin{center}
\begin{tabular}{|r|rrrrrrrrr|}\hline
 $i$&1&2&3&4&5&6&7&8&9\\\hline
 $m^{(22)}_{91i}$ & 0 & 0 & 4.1531 & -0.4627 & -0.3404 & -1.0326 & 0.0809 & 0.2167 & 0 \\[1mm]
 $m^{(22)}_{92i}$ & 0 & 0 & 12.4592 & 0.6940 & -1.7340 & 1.2360 & -0.1921 & 0.3998 & 0 \\[1mm]
 $m^{(22)}_{93i}$ & 0 & 0 & 0 & 0 & -7.6198 & -4.1683 & 1.5484 & 15.674 & 0 \\[1mm]
 $m^{(22)}_{94i}$ & 0 & 0 & 0 & 0 & -5.0848 & 3.2810 & 1.2592 & -1.2804 & 0 \\[1mm]
 $m^{(22)}_{95i}$ & 0 & 0 & 0 & 0 & -147.615 & -86.199 & 7.4486 & 106.285 & 0 \\[1mm]
 $m^{(22)}_{96i}$ & 0 & 0 & 0 & 0 & -67.9394 & 41.9523 & 0.2364 & 6.405 & 0 \\[1mm]
 $m^{(22)}_{99i}$ & 0 & 0 & 0 & 0 & 0 & 0 & 0 & 0 & 0 \\[1mm]
\hline
\end{tabular}
\caption{``Magic numbers'' $m^{(22)}_{9li}.$\label{tab:magic22}}
\end{center}
\end{table}

\end{appendix}

\section*{Acknowledgements}
We thank T. Hurth for many useful discussions and cross checks, and A. Bharucha, S. Descotes-Genon and J. Matias for useful comments.


\end{document}